\documentclass[aps, prc, 12pt, onecolumn, amsmath, amssymb, superscriptaddress, nofootinbib]{revtex4-2}
\usepackage{graphicx}
\usepackage{color}
\newcommand{\be}{\begin{equation}}
\newcommand{\ee}{\end{equation}}
\newcommand{\beq}{\begin{eqnarray}}
\newcommand{\eeq}{\end{eqnarray}}
\newcommand{\ba}{\begin{array}}
\newcommand{\ea}{\end{array}}
\begin{document}

\title{Universal Mass Equation for \\ Equal-Quantum Excited-States Sets~II}

\vspace{3mm}
\date{\today}

%--------------------- AUTHOR LIST --------------------------- 
\author{\mbox{L.~David~Roper}} 
\altaffiliation{Author: \texttt{roperld@vt.edu}} 
\affiliation{Prof. Emeritus of Physics, Virginia Polytechnic Institute and State University\\
    1001 Auburn Dr. SW, Blacksburg VA 24060, USA}

\author{\mbox{Igor~Strakovsky}} 
\altaffiliation{Author: \texttt{igor@gwu.edu}} 
\affiliation{Institute for Nuclear Studies, Department of Physics, 
    The George Washington University, Washington, DC 20052, USA}

%---------------------- ABSTRACT -----------------------------
\begin{abstract}
We extend our recent study of the universal mass equation for equal-quantum excited-states sets reported by Roper and Strakovsky~\cite{Roper:2024ovj}. The masses of twelve baryon sets and sixteen meson sets, with only two equal-quantum excited states in each set, using Breit-Wigner PDG2024 masses and their uncertainties at fixed $J^P$ for baryons and $J^{PC}$ for mesons, are fitted by a simple one-parameter logarithmic function, $M_n = \alpha Ln(n) + 
M_1$, where $n$ is the level of radial excitation. Two accurate masses that start a set are used to calculate four higher masses in the set accurately. It is noted that $\alpha$ values for  $b\bar{b}$ equal-quantum excited-states sets accurately lie on a straight line, 
 whose line parameters can be used to calculate $\alpha$ and predict higher mass states for $b\bar{b}$ sets that have only one known member.
 
 \end{abstract} 
\maketitle

%------------------------------------------------------------ 
%\clearpage 
\section{Introduction} 
\label{Sec:Intro}
\href{https://arxiv.org/pdf/2410.11196}{Roper and Strakovsky} (RS) \cite{Roper:2024ovj} have shown that the masses of baryon and meson equal-quantum excited-states sets reported in the \href{https://pdg.lbl.gov/2024/listings/contents_listings.html}{Particle Data Listings from the Particle Data Group (PDG) report}~\cite{ParticleDataGroup:2024cfk}  are well fitted by a simple equation, 
\begin{equation}
    \textbf{ $M_n$ = $\alpha$~Ln(n) + $\beta$}, \>
\label{eq:eq1}
\end{equation}
where $n$ is the level of radial excitation in the data set. Only state sets with three or more excited states were used. We use a Naming Scheme for Hadrons following Ref.~\cite{Amsler:2024}.

An excited state is a quantum state of a system that has a higher energy than the ground state $M_1$. In this work, the authors often label a ground state as an excited state of the vacuum.

This article reports using data sets with only two known excited states with masses M\textsubscript{1} and M\textsubscript{2} (duo sets), which have reasonably low measurement uncertainties. Here the parameter $\beta = M_1$.

It seems highly likely that, when only two excited states in a set have been measured, they are the first two states in a set. That may not be the case for all duo sets. There may be missing states in a few of the sets we study in this article, \textit{i.e.}, $M_1$ or $M_2$ may be missing, more likely $M_2$ than $M_1$. When that turns out to be the case later, the calculation for such a case needs to be redone.

It is shown below that, using only the two lowest excited states, masses of the next four higher states can be calculated with reasonable accuracy; for instance, for the N(1/2\textsuperscript{+}) baryon data set, which has six known members, and for the a\textsubscript{0} meson data set, which has four known members. (See Part I for calculations of N(1/2\textsuperscript{+}) baryon and a\textsubscript{0} meson.)

The authors conclude that excited-state data sets with only the two lowest masses (duo sets) known with reasonably high accuracy can be used to estimate, with mass uncertainties, the masses of the next four states in the set. Some duo sets have such small mass uncertainties that more than four higher-mass states can be accurately predicted; \textit{e.g.}, $s\bar{s}$, $c\bar{c}$, and $b\bar{b}$ states.

The excited-state masses and their uncertainties, in MeV units, used in this document are taken from \href{https://pdg.lbl.gov/2024/listings/contents_listings.html}{Particle Data Listings at PDG}~\cite{ParticleDataGroup:2024cfk}. 

%------------------------------------------------------------
%\clearpage
\section{Simple Mathematics}
The equations and propagated $\alpha$ and $\beta$ uncertainties for the two masses are:
\begin{equation}
    \textbf{$M_1$ = $\alpha$~Ln(1) + $\beta$ = $\beta$}~\text{and}~\delta{\beta} = \delta{M_1} \>,
    \label{eq:eq2}
\end{equation}
    \begin{equation}
    \textbf{$\alpha$ = ($M_2 - M_1$)/Ln(2)}~\text{and}~
    \delta{\alpha} = \sqrt{(\delta{M_1})^2+(\delta{M_2})^2} /Ln(2) \>.
    \label{eq:eq3}
\end{equation}
Predicted higher masses ($n>2$)  and their uncertainties are:
\begin{equation}
    \textbf {$M_n$ = $\alpha$~Ln(n) + $\beta$}~\text{and}~
    \delta{M_n} = \sqrt{[Ln(n)\delta{\alpha}]^2 + (\delta{\beta})^2} \>.
\label{eq:eq4}
\end{equation}

%---------------------------------------------------------------
\section{Test Cases}
\raggedright
Compare these 2-states calculation methods to two of the $\chi^2$-fitted data sets of RS~\cite{Roper:2024ovj}, $N1/2^+$ and $a_0$, which have several accurately measured masses. The purpose of the two test cases is to compare the two 2-state sets' third and higher predicted states' masses with measured data.

%-------------------------------------------------------------------
\subsection{$N1/2^+$ Baryon Data Set with Six Excited States (Test Case)}

\begin{figure}[htb!]
%\vspace{-0.3cm}
\centering
{
    \includegraphics[width=0.5\textwidth,keepaspectratio]{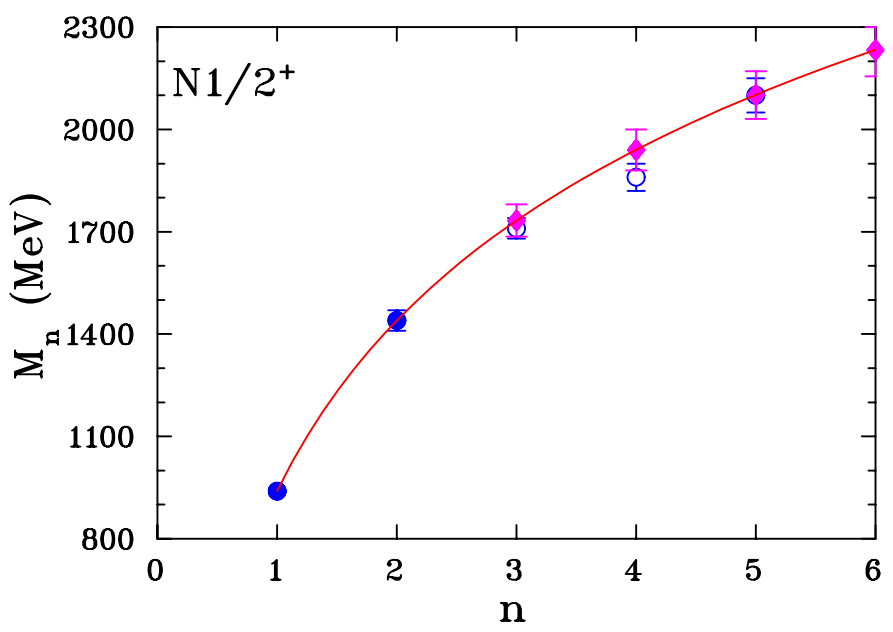} 
}

\centerline{\parbox{0.8\textwidth}{
\caption[] {\protect\small
Two $N1/2^+$ data used for the 2-states calculation are the two blue filled circles: $N(940)$ and $N(1440)$~\cite{ParticleDataGroup:2024cfk}. 
Three higher $N1/2^+$ data are shown by three blue open circles (see Table~\ref{tbl:N(1/2+)}).
Filled magenta diamonds are four 2-states predicted states. 
The solid red curve presents the best-fit result for the 2-states calculation.
}
\label{fig:N(1/2+)} } }
\end{figure}

See Table~\ref{tbl:N(1/2+)} and Fig.~\ref{fig:N(1/2+)} for the $N1/2^+$ measured data set and the calculation of two states at the measured data points. Note how well the 2-state calculation compares with the three higher measured data.
%--------------------------------------------
%--------------------------------------------
%---------------------------------------------------------
\begin{table}[htb!]

\centering \protect\caption{Two input excited-states' masses for the $N1/2^+$ data set 
and four 2-States calculated higher excited-states' masses. 
Data masses $m_3$, $m_4$, and $m_5$ are from PDG~\cite{ParticleDataGroup:2024cfk}. 
Parameter $\alpha$ is from RS~\cite{Roper:2024ovj} (right) 2-States (left).}
\vspace{2mm}
{%
\begin{tabular}{|c|c|c|c|c|}
\hline
Quantity  & 2-States &  Data   \tabularnewline
          &  (MeV)  & (MeV) \tabularnewline
\hline
$\alpha$  & 722.0$\pm$43.3& 698.2$\pm$14.1 \tabularnewline
m$_3$     & 1733$\pm$47   & 1710$\pm$30 \tabularnewline
m$_4$     & 1940$\pm$60   & 1860$\pm$40 \tabularnewline
m$_5$     & 2101$\pm$70   & 2100$\pm$50 \tabularnewline
m$_6$     & 2233$\pm$77   &  \tabularnewline
\hline
\end{tabular}} \label{tbl:N(1/2+)}
\end{table}

%-------------------------------------------------------
\subsection{$a_0$ Meson Data Set with Four Excited States (Test Case)}

\begin{figure}[htb!]
%\vspace{-0.3cm}
\centering
{
    \includegraphics[width=0.5\textwidth,keepaspectratio]{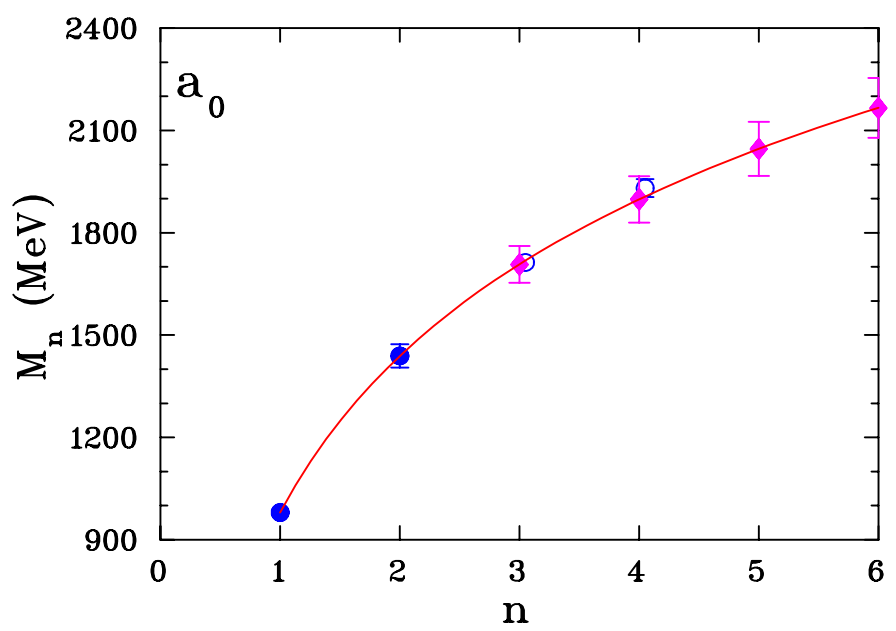} 
}

\centerline{\parbox{0.8\textwidth}{
\caption[] {\protect\small
Two $a_0$ data used for the 2-states calculation are the two blue filled circles: $a_0(980)$ and $a_0(1439)$~\cite{ParticleDataGroup:2024cfk}.
Two higher $a_0$ data are shown by blue open circles (see Table~\ref{tbl:a_0}).
Filled magenta diamonds are four 2-states predicted states. 
The solid red curve presents the best-fit result for the 2-states calculation.
}
\label{fig:a_0} } }
\end{figure}

See Table~\ref{tbl:a_0} and Fig.~\ref{fig:a_0} for the $a_0$ measured data set and the 2-state calculation using the two lowest measured data points. Note how well the 2-state calculation compares with the two higher measured data.
%--------------------------------------------
%--------------------------------------------
%---------------------------------------------------------
\begin{table}[htb!]

\centering \protect\caption{Two input excited-states' masses for the $a_0$ data set 
and four 2-States calculated higher excited-states' masses. 
Data masses $m_3$ and $m_4$ are from PDG~\cite{ParticleDataGroup:2024cfk}. 
Parameter $\alpha$ is from RS~\cite{Roper:2024ovj} (right)  and 2-States (left).}
\vspace{2mm}
{%
\begin{tabular}{|c|c|c|c|c|}
\hline
Quantity  & 2-State &  Data \tabularnewline
          &  (MeV)  & (MeV) \tabularnewline
\hline
$\alpha$  &662.2$\pm$49.0& 677.8$\pm$20.6 \tabularnewline
m$_3$     & 1707$\pm$54  & 1713$\pm$19    \tabularnewline
m$_4$     & 1898$\pm$68  & 1931$\pm$26    \tabularnewline
m$_5$     & 2046$\pm$79  &  \tabularnewline
m$_6$     & 2166$\pm$88  &  \tabularnewline
\hline
\end{tabular}} \label{tbl:a_0}
\end{table}

%------------------------------------------------------------
%\clearpage
\section{Baryon Two-Excited-States Sets}

%-----------------------------------------------------------------
%\subsection{Baryons ($J^P)$(12)} 

List of baryons studied: $\Delta 1/2^+$, $\Delta 3/2^-$, $\Delta 5/2^+$, $\Delta 5/2^-$, $\Delta 7/2^+$, $\Lambda 3/2^+$, $\Lambda 5/2^+$, $\Lambda 5/2^-$, $\Sigma5/2^+$, $\Lambda_c 3/2^-$, $\Xi_b 3/2^-$, and $P_{c\bar{c}s}^01/2^-$.

%---------------------------------------------------------------
\subsection{Delta Baryons}

%---------------------------------------------------------------
\subsubsection{$\Delta 1/2^+$ Two Known Excited States: Predict Four More}

Two excited states $\Delta 1/2^+$ are recorded in the Particle Data
Listings~\cite{ParticleDataGroup:2024cfk}. $\Delta 1/2^+$: $I(J^P)S = 3/2(1/2^+)0$. 

\begin{figure}[htb!]
%\vspace{-0.3cm}
\centering
{
    \includegraphics[width=0.5\textwidth,keepaspectratio]{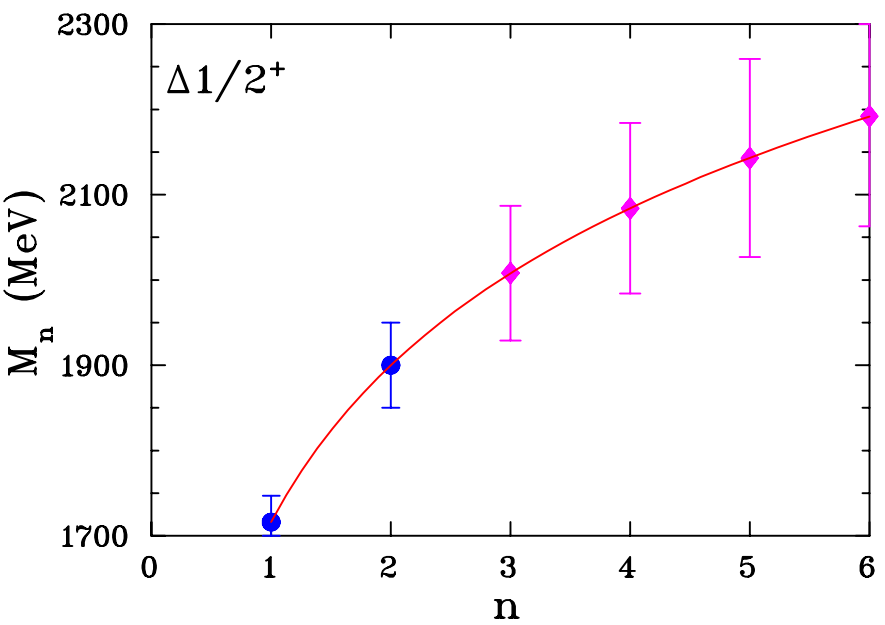} 
}

\centerline{\parbox{0.8\textwidth}{
\caption[] {\protect\small
Data for $\Delta 1/2^+$ (blue circles): $\Delta(1716)$ and $\Delta(1900)$~\cite{ParticleDataGroup:2024cfk}.
Predicted states (magenta diamonds): $\Delta(2008)$, $\Delta(2084)$, $\Delta(2143)$, and $\Delta(2192)$ with masses of $2008\pm 79~\mathrm{MeV}$, $2084\pm 100~\mathrm{MeV}$, 
$2143\pm 116~\mathrm{MeV}$, and $2192\pm 129~\mathrm{MeV}$, respectively.
The solid red curve presents the best-fit result for the 2-states calculation.
The fit parameter $\alpha = 265.5\pm 72.1~\mathrm{MeV}$.
}
\label{fig:D1} } }
\end{figure}

The logarithmic fit to the Breit-Wigner (BW) masses (MeV) of the two known excited states of $\Delta 1/2^+$ (blue circles) and the four higher excited states (magenta diamonds) projected is shown in Fig.~\ref{fig:D1}. 

%--------------------------------------------
%--------------------------------------------

%---------------------------------------------------------------
\subsubsection{$\Delta 3/2^-$ Two Known Excited States: Predict Four More}

Two excited states $\Delta 3/2^-$ are recorded in the Particle Data
Listings~\cite{ParticleDataGroup:2024cfk}. $\Delta 3/2^-$: $I(J^P)S = 3/2(3/2^-)0$. 

\begin{figure}[htb!]
%\vspace{-0.3cm}
\centering
{
    \includegraphics[width=0.5\textwidth,keepaspectratio]{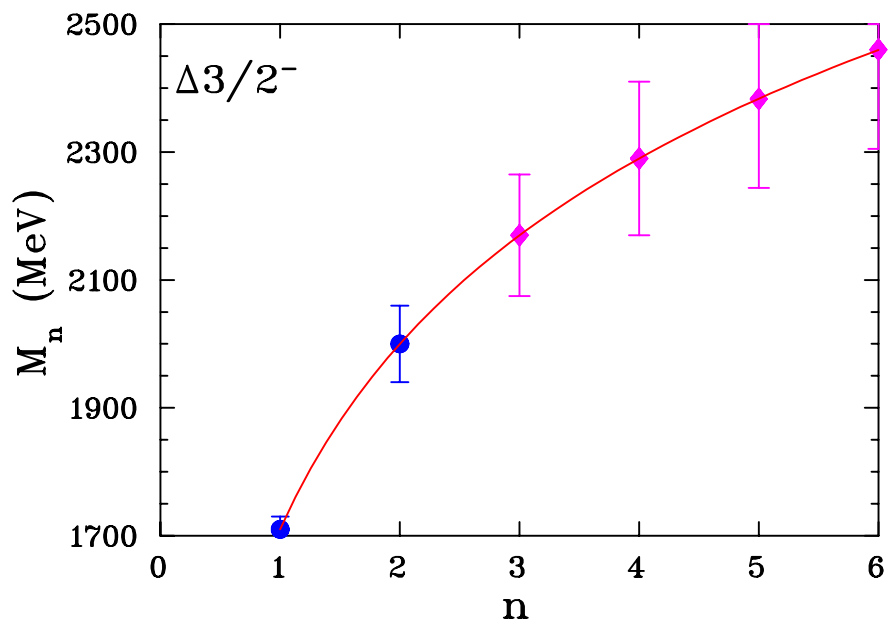} 
}

\centerline{\parbox{0.8\textwidth}{
\caption[] {\protect\small
Data for $\Delta 3/2^-$ (blue circles): $\Delta(1710)$ and $\Delta(2000)$~\cite{ParticleDataGroup:2024cfk}.
Predicted states (magenta diamonds): $\Delta(2170)$, $\Delta(2290)$, $\Delta(2383)$, and $\Delta(2460)$ with masses of $2170\pm 95~\mathrm{MeV}$, $2290\pm 120~\mathrm{MeV}$, 
$2383\pm 139~\mathrm{MeV}$, and $2460\pm 155~\mathrm{MeV}$, respectively.
The solid red curve presents the best-fit result for the 2-states calculation.
The fit parameter $\alpha = 418.4\pm 86.6~\mathrm{MeV}$.
}
\label{fig:D2} } }
\end{figure}

The logarithmic fit to the BW masses (MeV) of the two known excited states of $\Delta 3/2^-$ (blue circles) and four higher excited states (magenta diamonds) projected is shown in Fig.~\ref{fig:D2}.
%--------------------------------------------
%--------------------------------------------

%---------------------------------------------------------------
\subsubsection{$\Delta 5/2^+$ Two Known Excited States: Predict Four More}

Two excited states $\Delta 5/2^+$ are recorded in the Particle Data
Listings~\cite{ParticleDataGroup:2024cfk}. $\Delta 5/2^+$: $I(J^P)S = 3/2(5/2^+)0$. 

\begin{figure}[htb!]
%\vspace{-0.3cm}
\centering
{
    \includegraphics[width=0.5\textwidth,keepaspectratio]{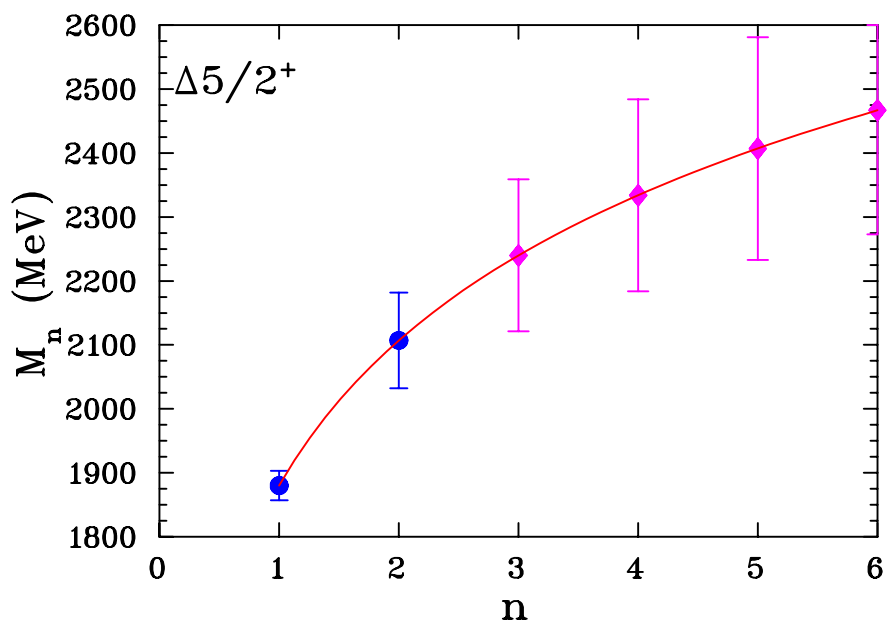} 
}

\centerline{\parbox{0.8\textwidth}{
\caption[] {\protect\small
Data for $\Delta 5/2^+$ (blue circles): $\Delta(1880)$ and $\Delta(2107)$~\cite{ParticleDataGroup:2024cfk}.
Predicted states (magenta diamonds): $\Delta(2240)$, $\Delta(2334)$, $\Delta(2407)$, and $\Delta(2467)$ with masses of $2240\pm 119~\mathrm{MeV}$, $2334\pm 150~\mathrm{MeV}$, 
$2407\pm 174~\mathrm{MeV}$, and $2467\pm 
194~\mathrm{MeV}$, respectively.
The solid red curve presents the best-fit result for the 2-states calculation.
The fit parameter $\alpha = 327.5\pm 108.2~\mathrm{MeV}$.
}
\label{fig:D3} } }
\end{figure}

The logarithmic fit to the BW masses (MeV) of the two known excited states of $\Delta 5/2^+$ (blue circles) and four higher excited states (red squares) projected is shown in Fig.~\ref{fig:D3}.
%--------------------------------------------
%--------------------------------------------

%---------------------------------------------------------------
\subsubsection{$\Delta 5/2^-$ Two Known Excited States: Predict Four More}

Two excited states $\Delta 5/2^-$ are recorded in the Particle Data
Listings~\cite{ParticleDataGroup:2024cfk}. $\Delta 5/2^-$: $I(J^P)S = 3/2(5/2^-)0$. 

\begin{figure}[htb!]
%\vspace{-0.3cm}
\centering
{
    \includegraphics[width=0.5\textwidth,keepaspectratio]{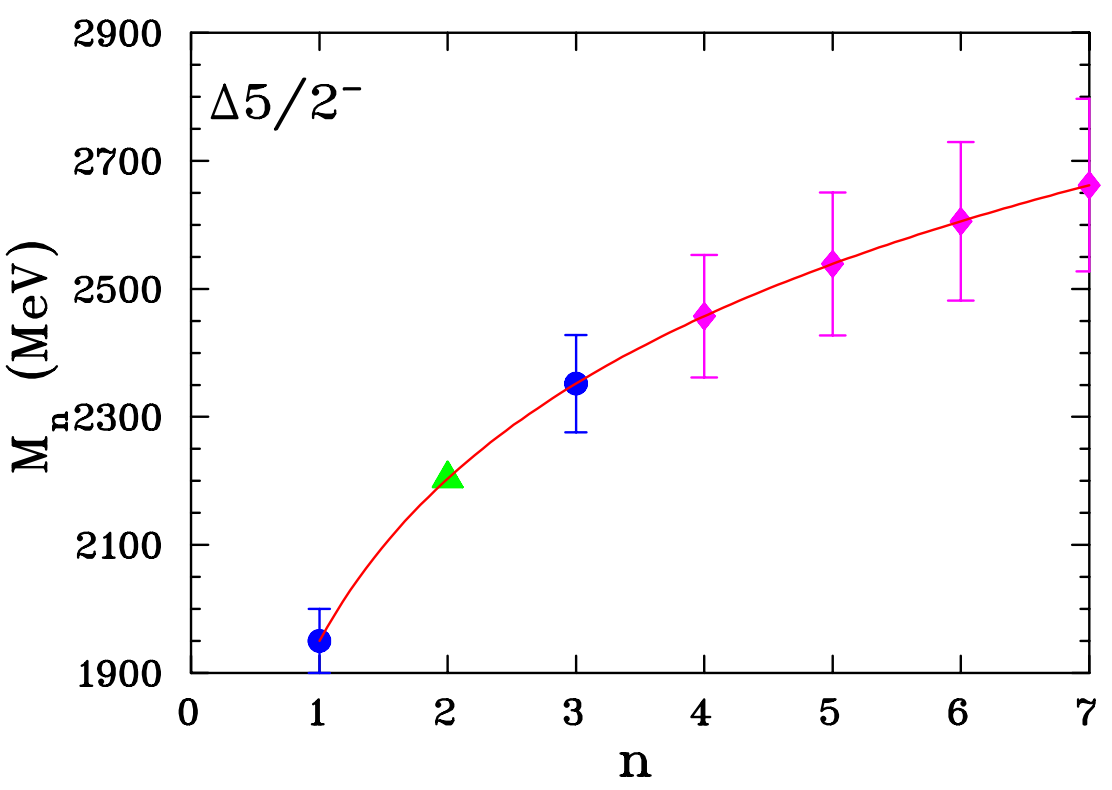} 
}

\centerline{\parbox{0.8\textwidth}{
\caption[] {\protect\small
Data for $\Delta 5/2^-$ (blue circles): $\Delta(1950)$ and $\Delta(2352)$~\cite{ParticleDataGroup:2024cfk}.
The green triangle is the calculated mass of the missing $N(2204)$ state with mass of $2204\pm 48~\mathrm{MeV}$.
Predicted states (magenta diamonds): $\Delta(2457)$, $\Delta(2539)$, $\Delta(2606)$, and $\Delta(2662)$ with masses of $2457\pm 96~\mathrm{MeV}$, $2539\pm 111~\mathrm{MeV}$, 
$2606\pm 124~\mathrm{MeV}$, and $2662\pm 135~\mathrm{MeV}$, respectively.
The solid red curve presents the best-fit result for the 2-states calculation.
The fit parameter $\alpha = 365.9\pm 69.2~\mathrm{MeV}$.
}
\label{fig:D4} } }
\end{figure}

The logarithmic fit to the BW masses (MeV) of the two known excited states of $\Delta 5/2^-$ (blue circles) and four higher excited states (red squares) projected is shown in Fig.~\ref{fig:D4}. In addition, a missing state (green triangle) is shown as calculated. (See section A. Baryon Power Equation as to why this missing state is included.)
%--------------------------------------------

%-------------------------------------------------------------------------------
\subsubsection{$\Delta 7/2^+$ Two Known Excited States: Predict Four More}

Two excited states $\Delta 7/2^+$ are recorded in the Particle Data
Listings~\cite{ParticleDataGroup:2024cfk}. $\Delta 7/2^+$: $I(J^P)S = 3/2(7/2^+)0$. 

\begin{figure}[htb!]
%\vspace{-0.3cm}
\centering
{
    \includegraphics[width=0.5\textwidth,keepaspectratio]{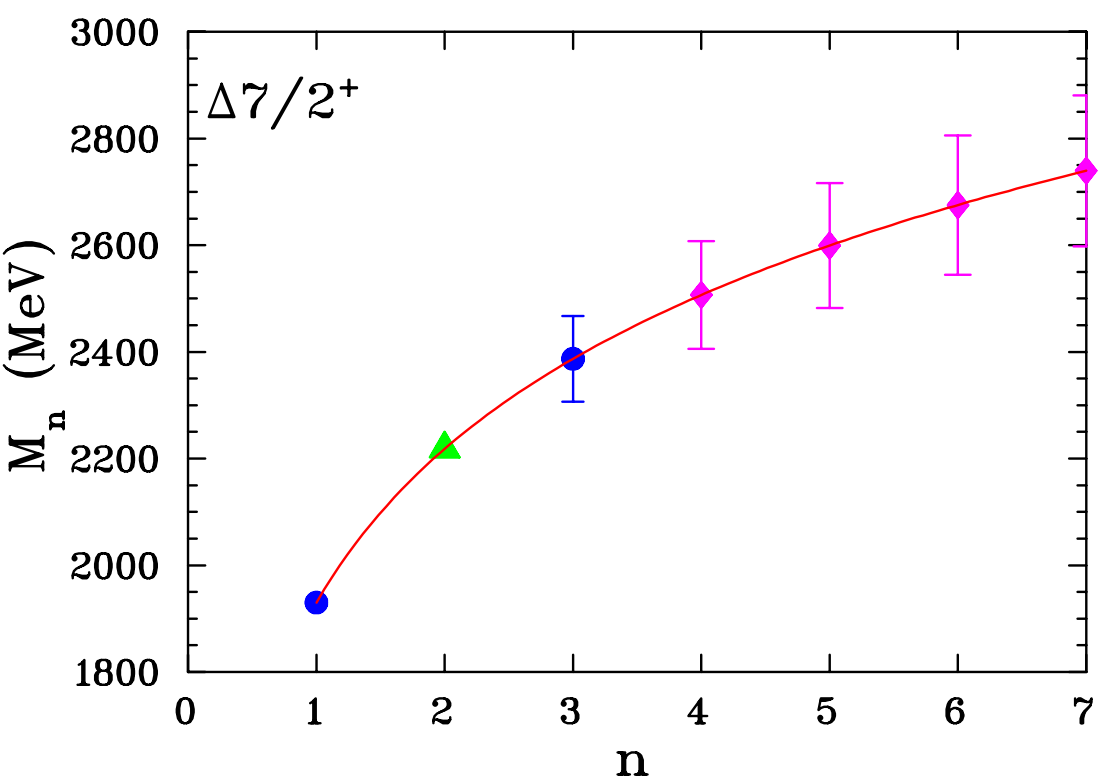} 
}

\centerline{\parbox{0.8\textwidth}{
\caption[] {\protect\small
Data for $\Delta 7/2^+$ (blue circles): $\Delta(1930)$ and 
$\Delta(2387)$~\cite{ParticleDataGroup:2024cfk}.
The green triangle is the calculated mass of the missing $N(2218)$ state with mass of $2218\pm 50~\mathrm{MeV}$.
Predicted states (magenta diamonds): $\Delta(2507)$, $\Delta(2599)$, $\Delta(2675)$, and $\Delta(2739)$ with masses of $2507\pm 101~\mathrm{MeV}$, $2599\pm 117~\mathrm{MeV}$, 
$2675\pm 130~\mathrm{MeV}$, and $2739\pm 142~\mathrm{MeV}$, respectively.
The solid red curve presents the best-fit result for the 2-states calculation.
The fit parameter $\alpha = 659.3\pm 115.4~\mathrm{MeV}$.
}
\label{fig:D5} } }
\end{figure}

The logarithmic fit to the BW masses (MeV) of the two known excited states of $\Delta 7/2^+$ (blue circles) and four higher excited states (red squares) projected is shown in Fig.~\ref{fig:D5}. In addition, a missing state (green triangle) is shown as calculated. (See section A. Baryon Power Equation as to why this missing state is included.)
%--------------------------------------------
%--------------------------------------------

%-------------------------------------------------------------------------------
\subsection{Lambda Baryons}

%-------------------------------------------------------------------------------
\subsubsection{$\Lambda 3/2^+$ Two Known Excited States: Predict Four More}

Two excited states $\Lambda 3/2^+$ are recorded in the Particle Data
Listings~\cite{ParticleDataGroup:2024cfk}. $\Lambda 3/2^+$: $I(J^P)S = 0(3/2^+)-1$. 

\begin{figure}[htb!]
%\vspace{-0.3cm}
\centering
{
    \includegraphics[width=0.5\textwidth,keepaspectratio]{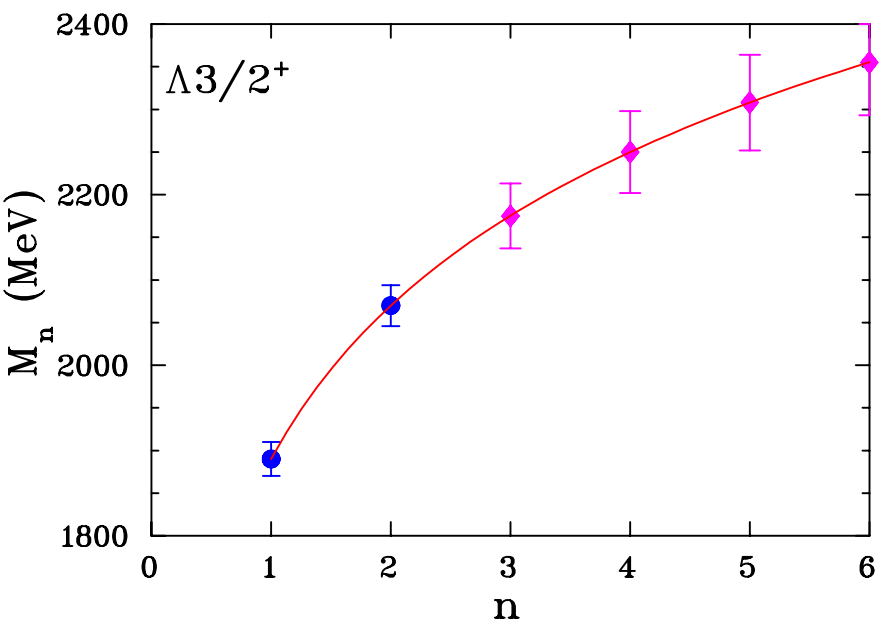} 
}

\centerline{\parbox{0.8\textwidth}{
\caption[] {\protect\small
Data for $\Lambda 3/2^+$ (blue circles): 
$\Lambda(1890)$ and 
$\Lambda(2070)$~\cite{ParticleDataGroup:2024cfk}.
Predicted states (magenta diamonds): $\Lambda(2175)$, $\Lambda(2250)$, $\Lambda(2308)$, and $\Lambda(2355)$ with masses of $2175\pm 38~\mathrm{MeV}$, $2250\pm 
48~\mathrm{MeV}$, 
$2308\pm 56~\mathrm{MeV}$, and $2355\pm 62~\mathrm{MeV}$, respectively.
The solid red curve presents the best-fit result for the 2-states calculation.
The fit parameter $\alpha = 259.7\pm 
34.6~\mathrm{MeV}$.
}
\label{fig:L1} } }
\end{figure}

The logarithmic fit to the BW masses (MeV) of the two known excited states of $\Lambda 3/2^+$ (blue circles) and four higher excited states (red squares) projected is shown in Fig.~\ref{fig:L1}.
%--------------------------------------------
%--------------------------------------------

%-------------------------------------------------------------------------------
\subsubsection{$\Lambda 5/2^+$ Two Known Excited States: Predict Four More}

Two excited states $\Lambda 5/2^+$ are recorded in the Particle Data
Listings~\cite{ParticleDataGroup:2024cfk}. $\Lambda 5/2^+$: $I(J^P)S = 0(5/2^+)-1$. 

\begin{figure}[htb!]
%\vspace{-0.3cm}
\centering
{
    \includegraphics[width=0.5\textwidth,keepaspectratio]{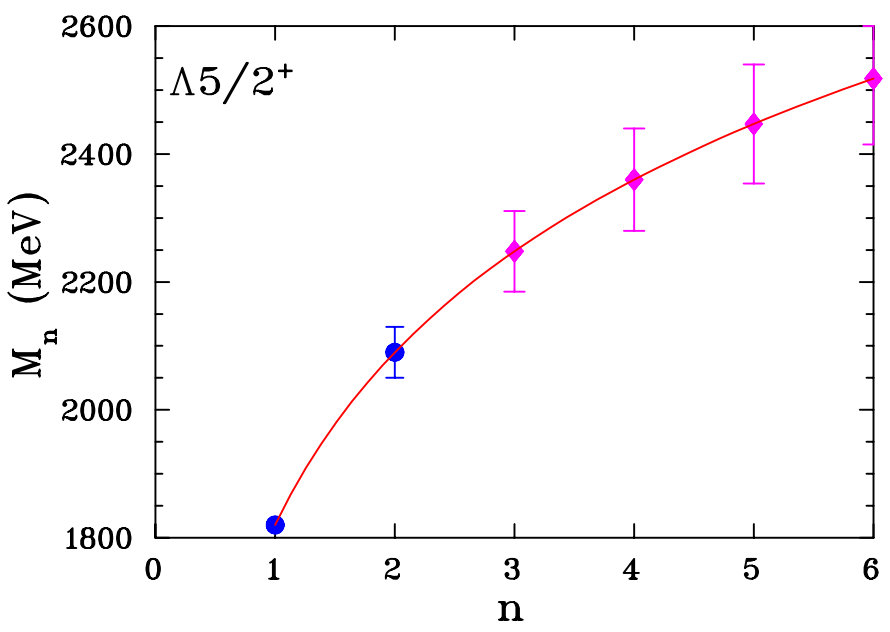} 
}

\centerline{\parbox{0.8\textwidth}{
\caption[] {\protect\small
Data for $\Lambda 5/2^+$ (blue circles): 
$\Lambda(1820)$ and 
$\Lambda(2090)$~\cite{ParticleDataGroup:2024cfk}.
Predicted states (magenta diamonds): $\Lambda(2248)$, $\Lambda(2360)$, $\Lambda(2447)$, and $\Lambda(2518)$ with masses of $2248\pm 63~\mathrm{MeV}$, $2360\pm 
80~\mathrm{MeV}$, $2447\pm 93~\mathrm{MeV}$, and $2518\pm 103~\mathrm{MeV}$.
The solid red curve presents the best-fit result for the 2-states calculation.
The fit parameter $\alpha = 389.5\pm 
57.7~\mathrm{MeV}$.
}
\label{fig:L2} } }
\end{figure}

The logarithmic fit to the BW masses (MeV) of the two known excited states of $\Lambda 5/2^+$ (blue circles) and four higher excited states (red squares) projected is shown in Fig.~\ref{fig:L2}.
%--------------------------------------------
%--------------------------------------------

%-------------------------------------------------------------------------------
\subsubsection{$\Lambda 5/2^-$ Two Known Excited States: Predict Four More}

Two excited states $\Lambda 5/2^-$ are recorded in the Particle Data
Listings~\cite{ParticleDataGroup:2024cfk}. $\Lambda 5/2^-$: $I(J^P)S = 0(5/2^-)-1$. 

\begin{figure}[htb!]
%\vspace{-0.3cm}
\centering
{
    \includegraphics[width=0.5\textwidth,keepaspectratio]{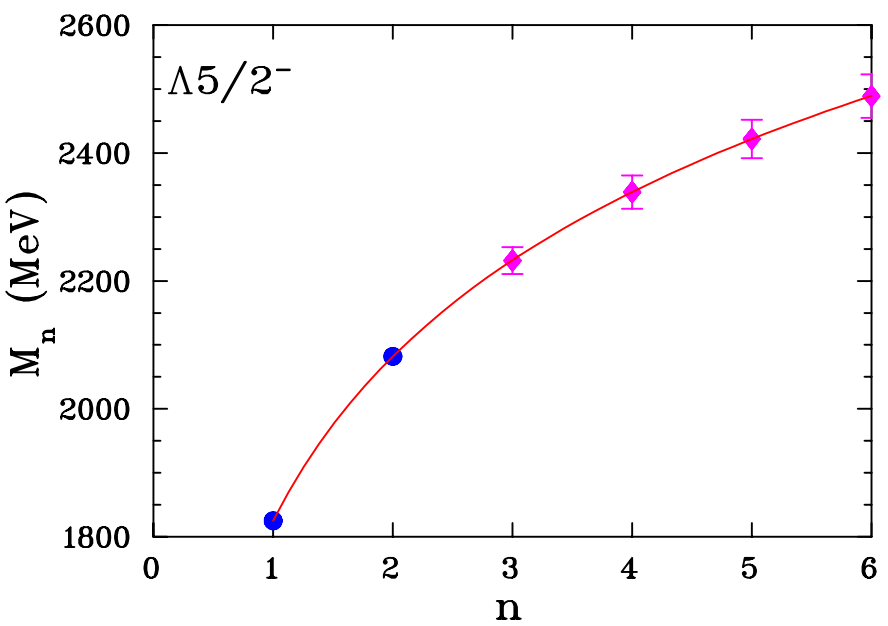} 
}

\centerline{\parbox{0.8\textwidth}{
\caption[] {\protect\small
Data for $\Lambda 5/2^-$ (blue circles): 
$\Lambda(1825)$ and 
$\Lambda(2082)$~\cite{ParticleDataGroup:2024cfk}.
Predicted states (magenta diamonds): $\Lambda(2232)$, $\Lambda(2339)$, $\Lambda(2422)$, and $\Lambda(2489)$ with masses of $2232\pm 21~\mathrm{MeV}$, $2339\pm 
26~\mathrm{MeV}$, $2422\pm 30~\mathrm{MeV}$, and $2489\pm 34~\mathrm{MeV}$. (Note that the uncertainties are so small that many more states can be predicted to high accuracy.)
The solid red curve presents the best-fit result for the 2-states calculation.
The fit parameter $\alpha = 370.8\pm 
18.7~\mathrm{MeV}$.
}
\label{fig:L3} } }
\end{figure}

The logarithmic fit to the BW masses (MeV) of the two known excited states of $\Lambda 5/2^-$ (blue circles) and four higher excited states (red squares) projected is shown in Fig.~\ref{fig:L3}.
%--------------------------------------------
%--------------------------------------------

%---------------------------------------------------------------------
\subsection{Sigma Baryons}

%----------------------------------------------------------------------
\subsubsection{$\Sigma 5/2^+$ Two Known Excited States: Predict Four More}

Two excited states $\Sigma 5/2^+$ are recorded in the Particle Data
Listings~\cite{ParticleDataGroup:2024cfk}. $\Sigma 5/2^+$: $I(J^P)S = 1(5/2^+)-1$. 

\begin{figure}[htb!]
%\vspace{-0.3cm}
\centering
{
    \includegraphics[width=0.5\textwidth,keepaspectratio]{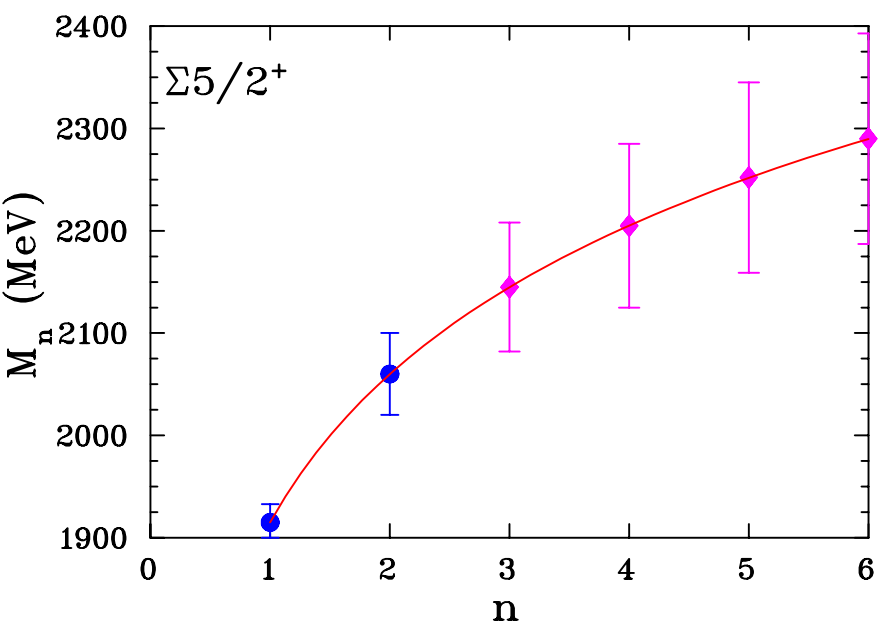} 
}

\centerline{\parbox{0.8\textwidth}{
\caption[] {\protect\small
Data for $\Sigma 5/2^+$ (blue circles): 
$\Sigma(1915)$ and 
$\Sigma(2060)$~\cite{ParticleDataGroup:2024cfk}.
Predicted states (magenta diamonds): $\Sigma(2145)$, $\Sigma(2205)$, $\Sigma(2252)$, and $\Sigma(2290)$ with masses of $2145\pm 63~\mathrm{MeV}$, $2205\pm 
80~\mathrm{MeV}$, $2252\pm 93~\mathrm{MeV}$, and $2290\pm 103~\mathrm{MeV}$.
The solid red curve presents the best-fit result for the 2-states calculation.
The fit parameter $\alpha = 209.2\pm 
57.7~\mathrm{MeV}$.
}
\label{fig:S1} } }
\end{figure}

The logarithmic fit to the BW masses (MeV) of the two known excited states of $\Sigma 5/2^+$ (blue circles) and four higher excited states (red squares) projected is shown in Fig.~\ref{fig:S1}.
%--------------------------------------------
%--------------------------------------------

%--------------------------------------------------------------------
\subsection{Charmed Baryons}

%-------------------------------------------------------------------
\subsubsection{$\Lambda_c3/2^-$ Two Known Excited States: Predict Four More}

Two $\Lambda_c3/2^-$ pion-nucleon scattering resonances are recorded in the Particle Data
Listings~\cite{ParticleDataGroup:2024cfk}. $\Lambda_c3/2^-$: $I(J^P)S,C = 0(3/2^-)-1,1$. 

\begin{figure}[htb!]
%\vspace{-0.3cm}
\centering
{
    \includegraphics[width=0.5\textwidth,keepaspectratio]{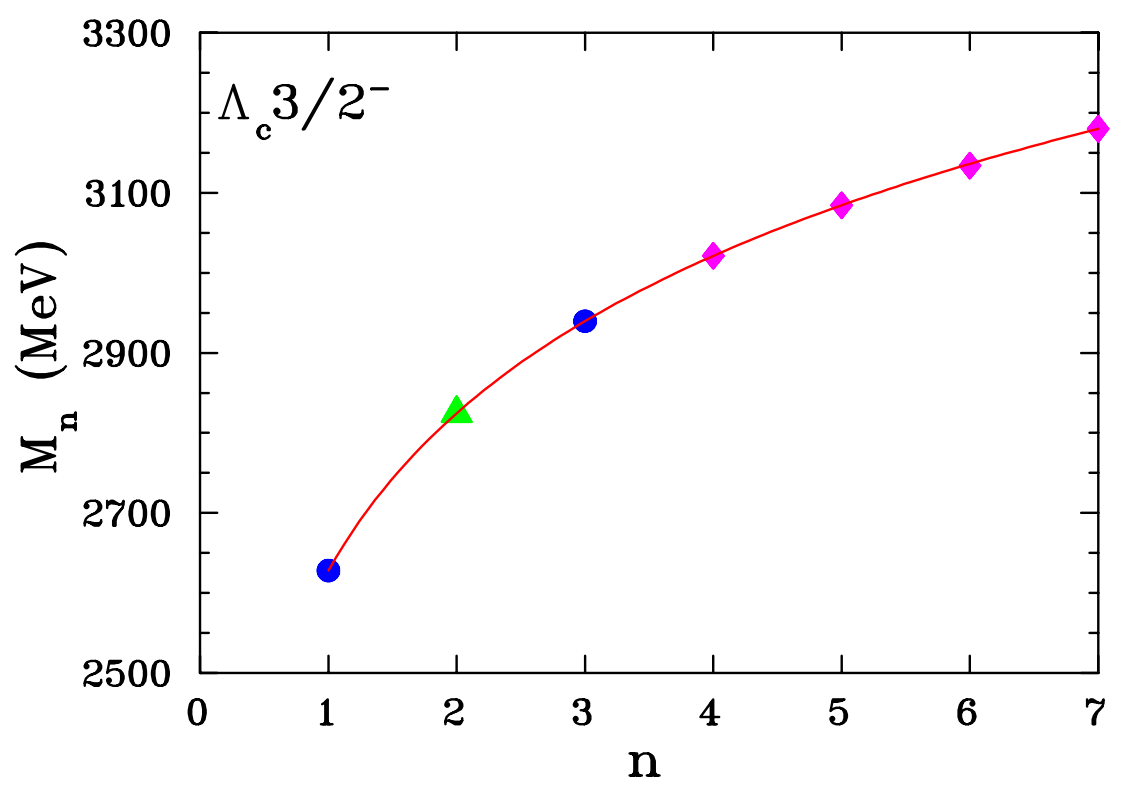} 
}

\centerline{\parbox{0.8\textwidth}{
\caption[] {\protect\small
Data for $\Lambda_c3/2^-$ (blue circles): 
$\Lambda_c(2628)$ and 
$\Lambda_c(2940)$~\cite{ParticleDataGroup:2024cfk}.
The green triangle is the calculated mass of the missing $N(2825)$ state with mass of $2824.6\pm 0.9~\mathrm{MeV}$.
Predicted states (magenta diamonds): $\Lambda_c(3021)$, $\Lambda_c(3085)$, $\Lambda_c(3134)$, and $\Lambda_c(3180)$ with masses of $3021.2\pm 1.8~\mathrm{MeV}$, $3084.5\pm 
2.1~\mathrm{MeV}$, $3133.9\pm 2.3~\mathrm{MeV}$, and $3179.9\pm 2.5~\mathrm{MeV}$. (Note that the uncertainties are so small that many more states can be predicted to high accuracy.)
The solid red curve presents the best-fit result for the 2-states calculation.
The fit parameter $\alpha = 449.5\pm 
2.0~\mathrm{MeV}$.
}
\label{fig:Lc} } }
\end{figure}

The logarithmic fit to the BW masses (MeV) of the two known excited states of $\Lambda_c3/2^-$ (blue circles) and four higher excited states (red squares) projected is shown in Fig.~\ref{fig:Lc}. In addition, a missing state (green triangle) is shown as calculated. (See section A. Baryon Power Equation as to why this missing state is included.)
%--------------------------------------------
%--------------------------------------------

%-------------------------------------------------------------------------------
\subsection{Bottom Baryons}

%-------------------------------------------------------------------------------
\subsubsection{$\Xi_b3/2^-$ Two Known Excited States: Predict Four More}

Two $\Xi_b3/2^-$ pion-nucleon scattering resonances are recorded in the Particle Data
Listings~\cite{ParticleDataGroup:2024cfk}. $\Xi_b3/2^-$: $I(J^P)S,C = 1/2(3/2^-)-2,1$. 

\begin{figure}[htb!]
%\vspace{-0.3cm}
\centering
{
    \includegraphics[width=0.5\textwidth,keepaspectratio]{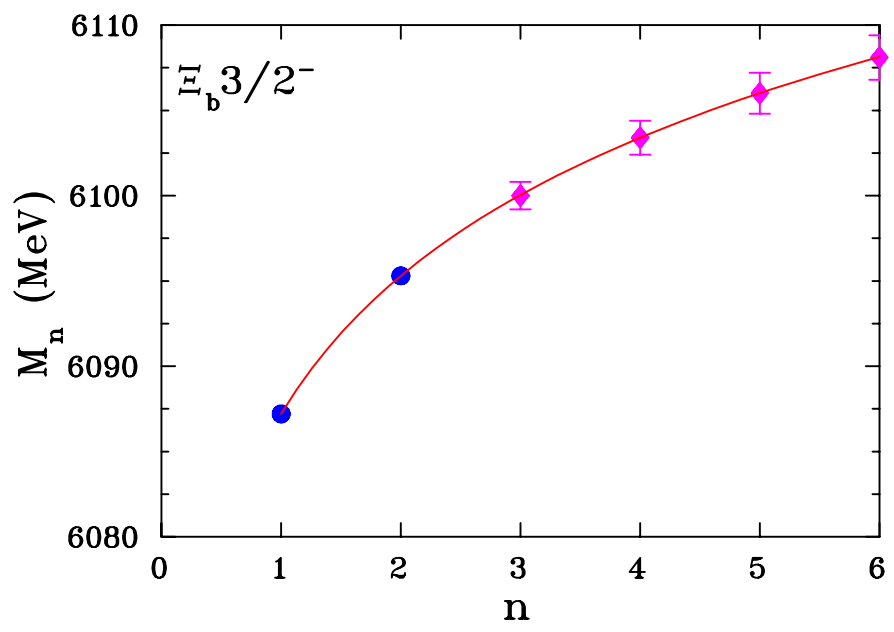} 
}

\centerline{\parbox{0.8\textwidth}{
\caption[] {\protect\small
Data for $\Xi_b3/2^-$ (blue circles): 
$\Xi_b(6087)$ and 
$\Xi_b(6095)$~\cite{ParticleDataGroup:2024cfk}.
Predicted states (magenta diamonds): $\Xi_b(6100)$, $\Xi_b(6103)$, $\Xi_b(6106)$, and $\Xi_b(6108)$ with masses of $6100.0\pm 0.8~\mathrm{MeV}$, $6103.4\pm 
1.0~\mathrm{MeV}$, $6106.0\pm 1.2~\mathrm{MeV}$, and $6108.1\pm 1.3~\mathrm{MeV}$.
(Note that the uncertainties are so small that many more states can be predicted to
high accuracy.) The solid red curve presents the best-fit result for the 2-states calculation.
The fit parameter $\alpha = 11.7\pm 
0.7~\mathrm{MeV}$.
}
\label{fig:Xb} } }
\end{figure}

The logarithmic fit to the BW masses (MeV) of the two known excited states of $\Xi_b3/2^-$ (blue circles) and four higher excited states (red squares) projected is shown in Fig.~\ref{fig:Xb}.
%--------------------------------------------
%--------------------------------------------

%-------------------------------------------------------------------------------
\subsection{Exotic Baryons}

%-------------------------------------------------------------------------------
\subsubsection{$P_{c\bar{c}s}^01/2^-$ Two Known Excited States: Predict Four More}

Two $P_{c\bar{c}s}^0$ pion-nucleon scattering resonances are recorded in the Particle Data
Listings~\cite{ParticleDataGroup:2024cfk}. $P_{c\bar{c}s}^0$: $I(J^P) = 0(1/2^-)$. 

\begin{figure}[htb!]
\vspace{-0.3cm}
\centering
{
   \includegraphics[width=0.5\textwidth,keepaspectratio]{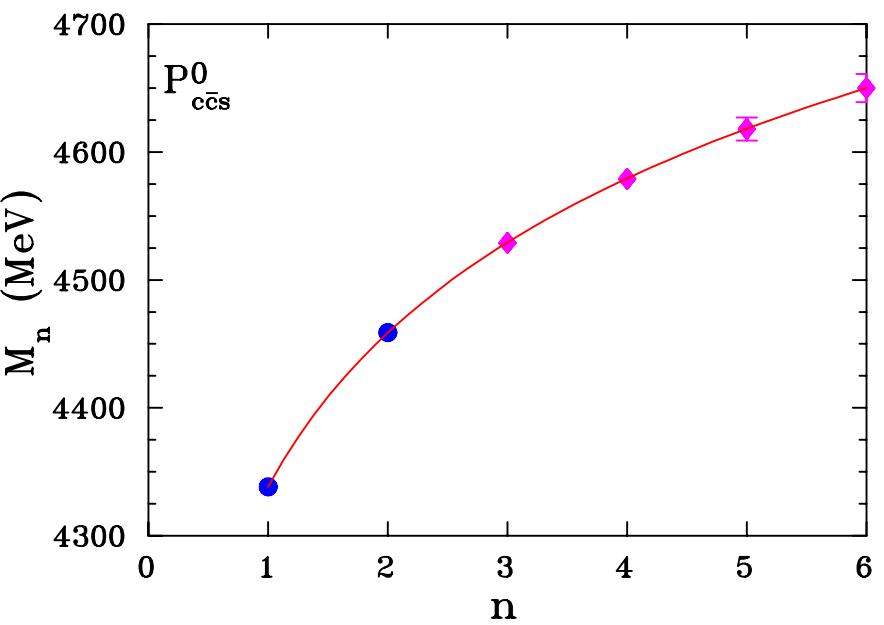} 
}

\centerline{\parbox{0.8\textwidth}{
\caption[] {\protect\small
Data for $P_{c\bar{c}s}^01/2^-$ (blue circles): 
$P_{c\bar{c}s}(4338)^0$ and 
$P_{c\bar{c}s}(4459)^0$~\cite{ParticleDataGroup:2024cfk}.
Predicted states (magenta diamonds): $P_{c\bar{c}s}(4529)^0$, $P_{c\bar{c}s}(4579)^0$, 
$P_{c\bar{c}s}(4618)^0$, and $P_{c\bar{c}s}(4650)^0$ 
with masses of $4529\pm 6~\mathrm{MeV}$, $4579\pm 
8~\mathrm{MeV}$, $4618\pm 9~\mathrm{MeV}$, and $4650\pm 
11~\mathrm{MeV}$. (Note that the uncertainties are so small that many more states can be predicted to high accuracy.)
The solid red curve presents the best-fit result for the 2-states calculation.
The fit parameter $\alpha = 174.0\pm 
5.9~\mathrm{MeV}$.
}
\label{fig:Pc} } }
\end{figure}

The logarithmic fit to the BW masses (MeV) of the two known excited states of $P_{c\bar{c}s}^0$ (blue 
circles) and four higher excited states (red squares) projected is shown in Fig.~\ref{fig:Pc}.
%--------------------------------------------
%--------------------------------------------

%-------------------------------------------------------------------------------
\subsection{Cumulative Baryon Excited States}

\begin{figure}[htb!]
%\vspace{-0.3cm}
\centering
{
    \includegraphics[width=0.43\textwidth,keepaspectratio]{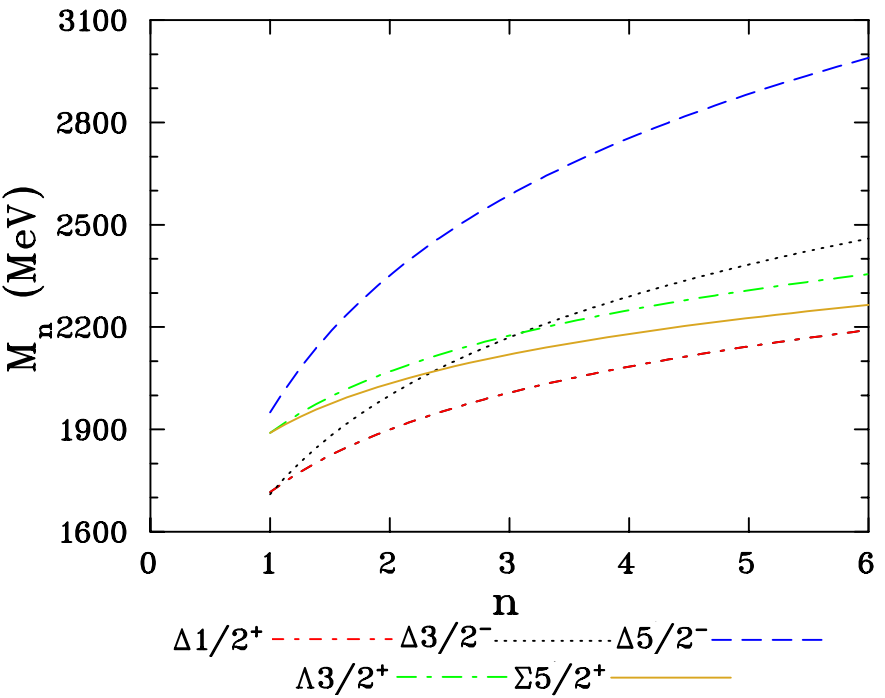}~~~ 
    \includegraphics[width=0.43\textwidth,keepaspectratio]{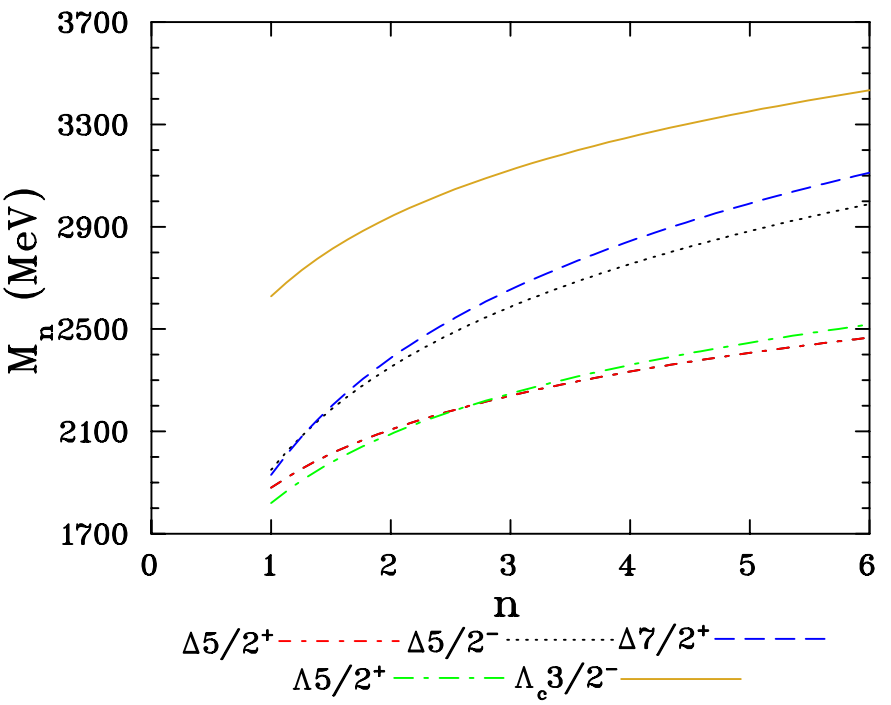}
    \includegraphics[width=0.43\textwidth,keepaspectratio]{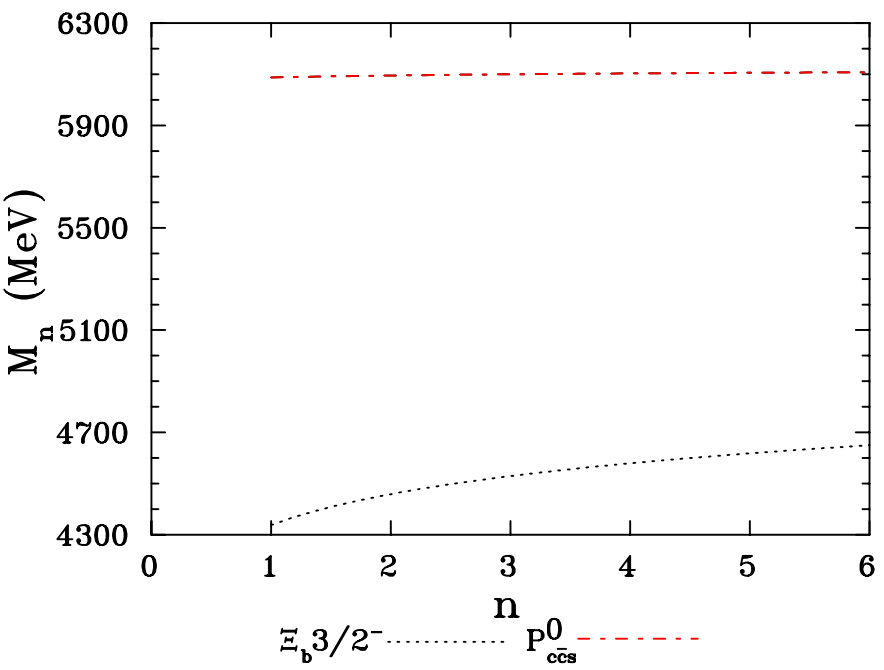}
}

\centerline{\parbox{0.8\textwidth}{
\caption[] {\protect\small
Logarithmic curves for twelfth baryon data sets that only have two known excited states.
\underline{Top Left}: $\Delta 1/2^+$, $\Delta 3/2^-$, $\Lambda 3/2^+$, $\Lambda 5/2^-$, and $\Sigma 5/2^+$.
\underline{Top Right}: $\Delta 5/2^+$, $\Delta 5/2^-$, $\Delta 7/2^+$, $\Lambda 5/2^+$, $\Lambda_c 3/2^-$, and $\Xi_b 3/2^-$.   
\underline{Bottom}: $P_{c\bar{c}s}^01/2^-$ and $\Xi_b 3/2^-$. 
}
\label{fig:meg} } }
\end{figure}

The cumulative mass curves of twelve baryon sets with equal-quantum excited states with only two known excited states per set are shown in Fig.~\ref{fig:meg}. The logarithmic slope, $\alpha$, usually decreases as the mass of the ground state increases.

\begin{figure}[htb!]
%\vspace{-0.3cm}
\centering
{
    \includegraphics[width=0.45\textwidth,keepaspectratio]{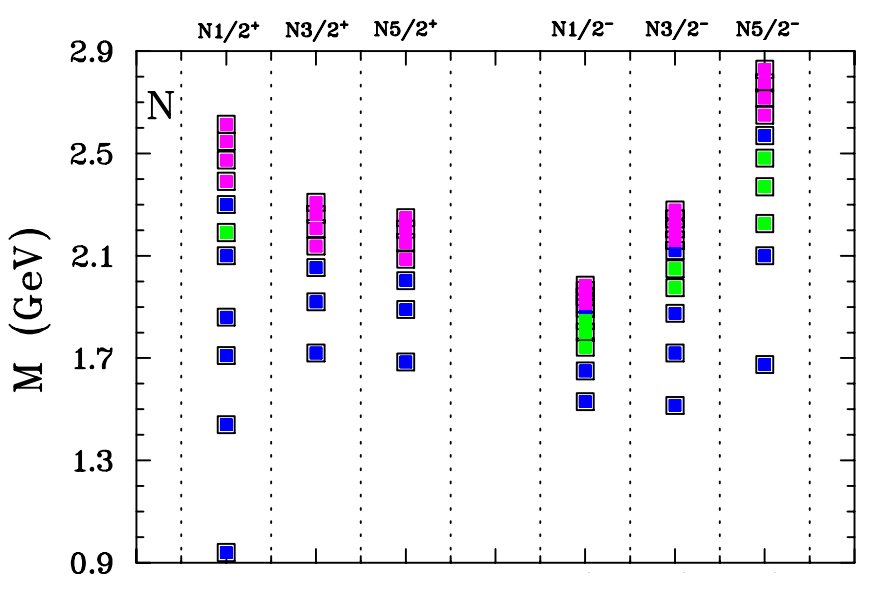}~~~ 
    \includegraphics[width=0.45\textwidth,keepaspectratio]{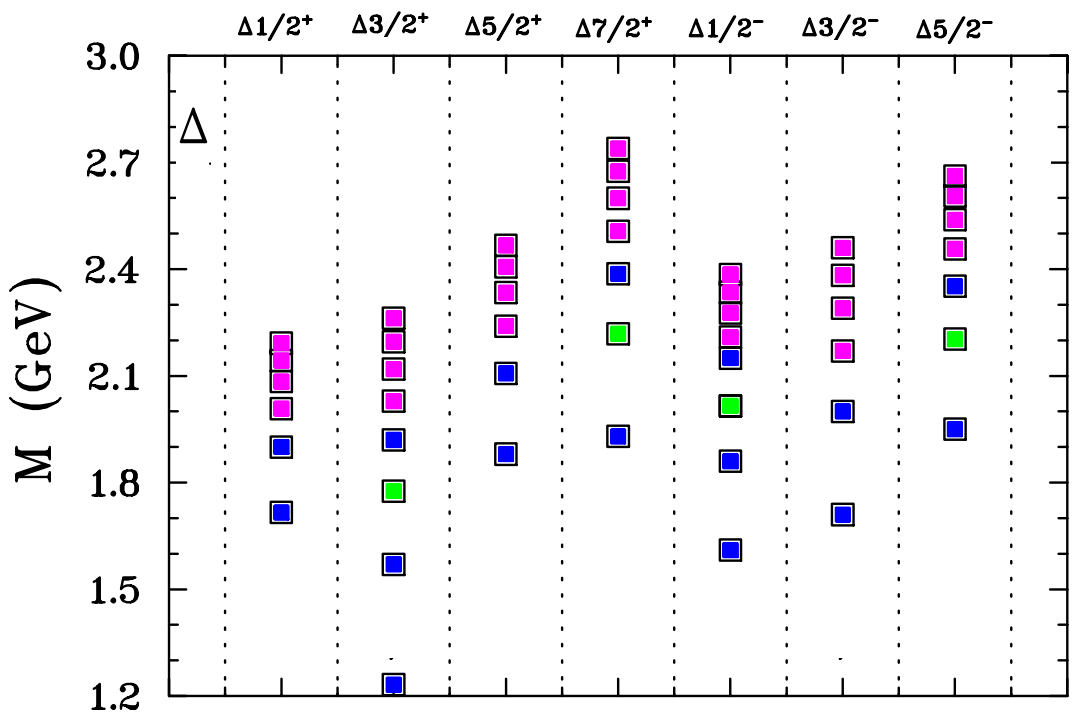}
    \includegraphics[width=0.45\textwidth,keepaspectratio]{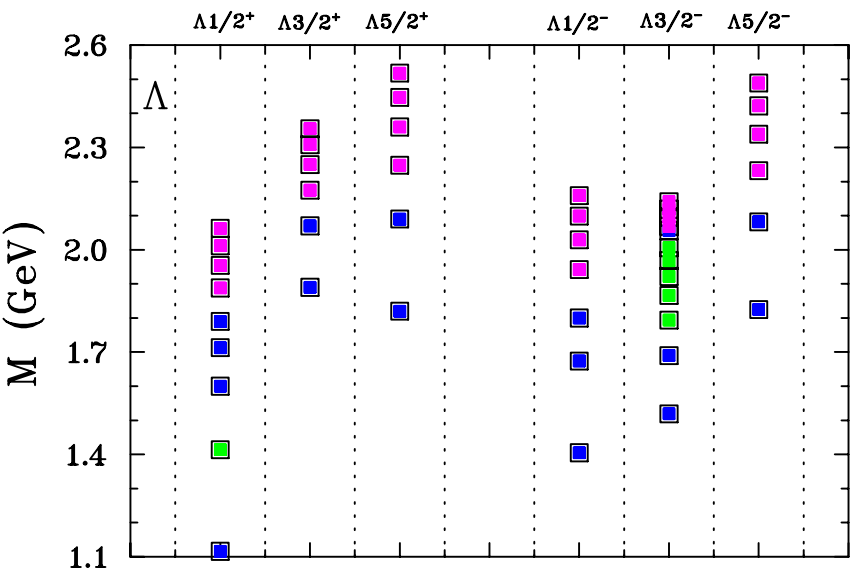}~~~ 
    \includegraphics[width=0.45\textwidth,keepaspectratio]{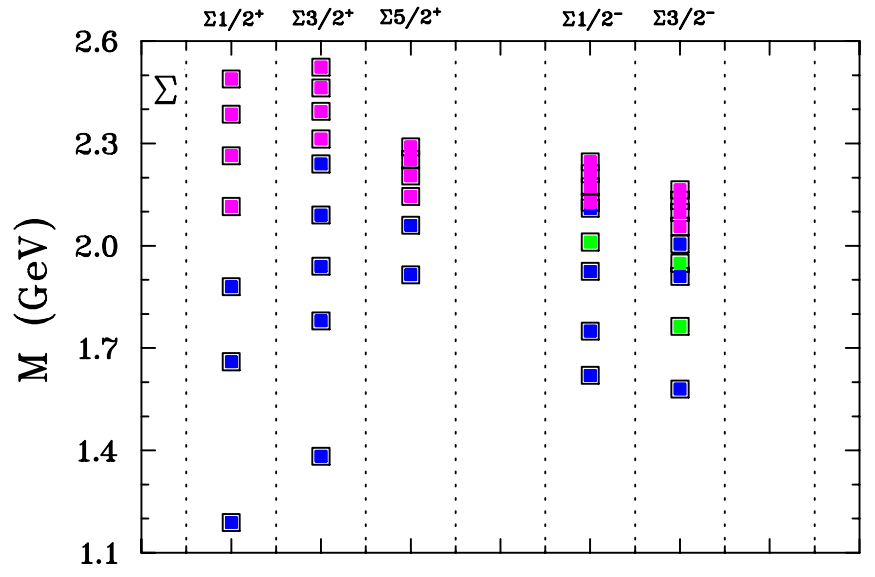} 
}

\centerline{\parbox{0.8\textwidth}{
\caption[] {\protect\small
Samples of spectral diagrams of baryon excited-states sets analyzed are shown versus $J^P$. Colors are used to display as follows: blue for PDG values~\cite{ParticleDataGroup:2024cfk} ($n = 3+$ states came from Part~I~\cite{Roper:2024ovj} and $n = 2$ states came from the current paper), green for missed states, and magenta for predicted states.
}
\label{fig:Money1} } }
\end{figure}

The analyzed spectra of the $N$, $\Delta$, $\Lambda$, and $\Sigma$ families of baryons for spins up to $5/2$ and both parities are shown in Figure~\ref{fig:Money1}.
%--------------------------------------------
%--------------------------------------------
%--------------------------------------------
%------------------------------------------------------

%----------------------------------------------------------------
\section{Meson Two-Excited-States}

%------------------------------------------------------------
%\clearpage
%\subsection{Mesons (I,JP)(9)}

List of mesons studied: $a_1(1^{++})$, $a_2(2^{++})$, $\eta_2(2^{-+})$, $f_4(4^{++})$, 
$K^\ast_2(2^+)$, 
%$K^\ast_4(4^+)$, 
$D^\ast_1(1^-)$, $D_{s1}(1^+)$, $D^\ast_{s1}(1^-)$, 
$B_c(0^-)$, $\eta_c(0^{-+})$, $\chi_{c2}(2^{++})$, $\eta_b(0^-)$, $h_b(1^{+-})$, $\chi_{b0}(0^{++})$, $T_{c\bar{c}\bar{s}1}(1^+)$, and $T_{b\bar{b}1}(1^{+-})$.

%---------------------------------------------------------------
\subsection{Light Unflavored Mesons}

%---------------------------------------------------------------
\subsubsection{$a_1(1^{++})$ Two Known Excited States: Predict Four More}

Two excited states $a_1(1^{++})$ are recorded in the Particle Data
Listings~\cite{ParticleDataGroup:2024cfk}. $a_1(1^{++})$: $I^G~(J^{PC})S = 1^-(1^{++})0$.

\begin{figure}[htb!]
%\vspace{-0.3cm}
\centering
{
    \includegraphics[width=0.5\textwidth,keepaspectratio]{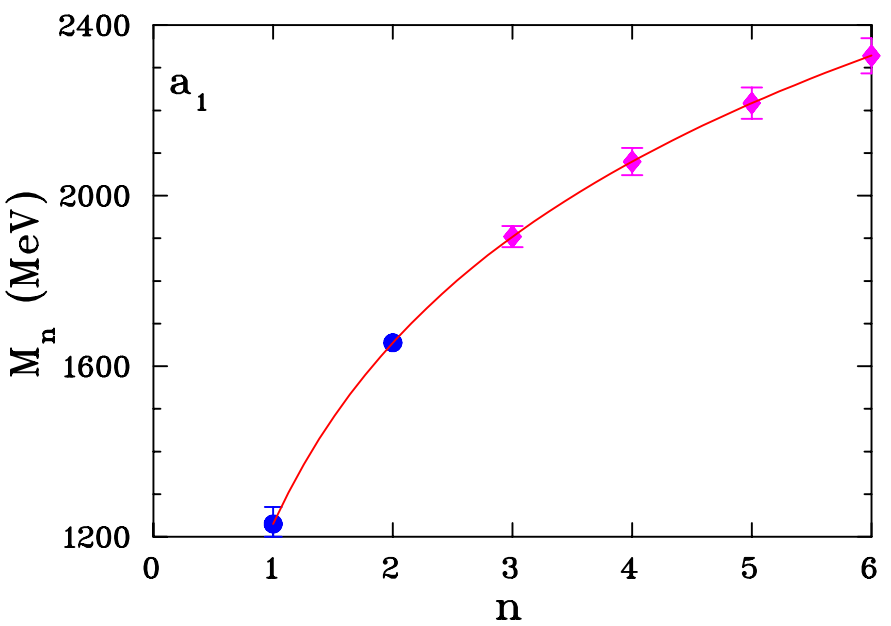} 
}

\centerline{\parbox{0.8\textwidth}{
\caption[] {\protect\small
Data for $a_1(1^{++})$ (blue circles): 
$a_1(1230)$ and 
$a_1(1655)$~\cite{ParticleDataGroup:2024cfk}.
Predicted states (magenta diamonds): $a_1(1904)$, $a_1(2080)$, $a_1(2217)$, and $a_1(2329)$ with masses of $1904\pm 25~\mathrm{MeV}$, $2080\pm 
32~\mathrm{MeV}$, $2217\pm 37~\mathrm{MeV}$, and $2329\pm 41~\mathrm{MeV}$.
The solid red curve presents the best-fit result for the 2-states calculation.
The fit parameter $\alpha = 613.1\pm 
23.1~\mathrm{MeV}$.
}
\label{fig:a1} } }
\end{figure}

The logarithmic fit to the BW masses (MeV) of the two known excited states of $a_1(1^{++})$ (blue circles) and four higher excited states (magenta diamonds) projected is shown in Fig.~\ref{fig:a1}.
%--------------------------------------------
%--------------------------------------------

%-------------------------------------------------------------------------------
\subsubsection{$a_2(2^{++})$ Two Known Excited States: Predict Four More}

Two excited states $a_2(2^{++})$ are recorded in the Particle Data
Listings~\cite{ParticleDataGroup:2024cfk}. $a_2(2^{++})$: $I^G~(J^{PC})S = 1^-(2^{++})0$. 

\begin{figure}[htb!]
%\vspace{-0.3cm}
\centering
{
    \includegraphics[width=0.5\textwidth,keepaspectratio]{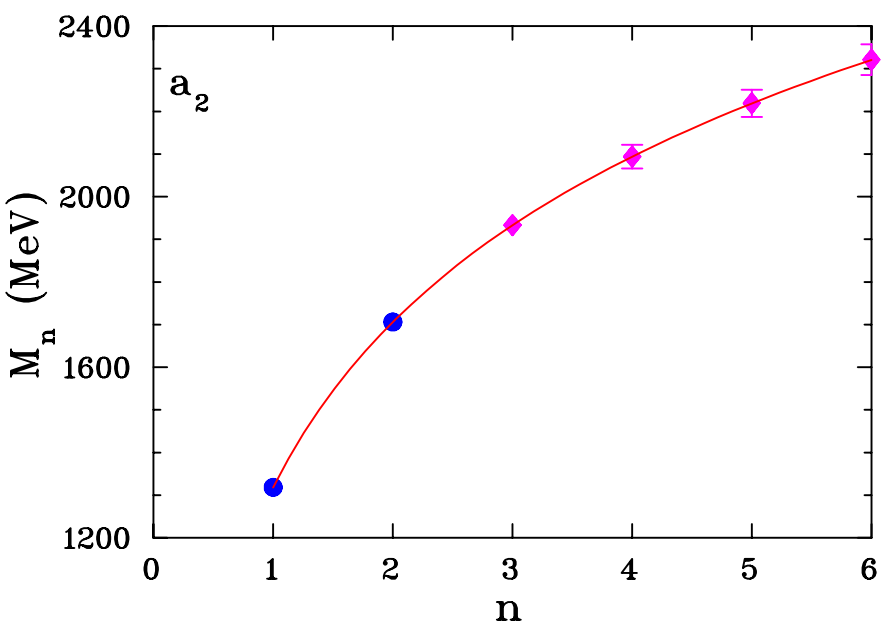} 
}

\centerline{\parbox{0.8\textwidth}{
\caption[] {\protect\small
Data for $a_2(2^{++})$ (blue circles): $a_2(1318)$ and 
$a_2(1706)$~\cite{ParticleDataGroup:2024cfk}.
Predicted states (magenta diamonds): $a_2(1933)$, $a_2(2094)$, $a_2(2219)$, and $a_2(2321)$ with masses of $1933\pm 22~\mathrm{MeV}$, $2094\pm 
28~\mathrm{MeV}$, $2219\pm 32~\mathrm{MeV}$, and $2321\pm 36~\mathrm{MeV}$. (Note that the uncertainties are so small that many more states can be predicted to high accuracy.)
The solid red curve presents the best-fit result for the 2-states calculation.
The fit parameter $\alpha = 559.5\pm 
20.2~\mathrm{MeV}$.
}
\label{fig:a2} } }
\end{figure}

The logarithmic fit to the BW masses (MeV) of the two known excited states of $a_2(2^{++})$ (blue circles) and four higher excited states (magenta diamonds) projected is shown in Fig.~\ref{fig:a2}.
%--------------------------------------------
%--------------------------------------------

%-------------------------------------------------------------------------------
\subsubsection{$\eta_2(2^{-+})$ Two Known Excited States: Predict Four More}

Two excited states $\eta_2(2^{-+})$ are recorded in the Particle Data
Listings~\cite{ParticleDataGroup:2024cfk}. $\eta_2(2^{-+})$: $I^G~(J^{PC})S = 0^+(2^{-+})0$.

\begin{figure}[htb!]
%\vspace{-0.3cm}
\centering
{
    \includegraphics[width=0.5\textwidth,keepaspectratio]{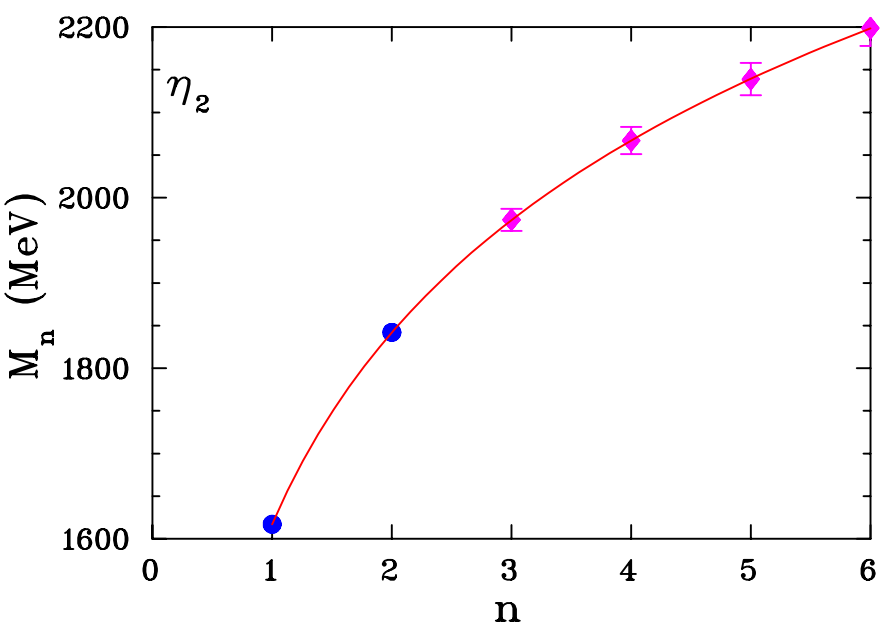} 
}

\centerline{\parbox{0.8\textwidth}{
\caption[] {\protect\small
Data for $\eta_2(2^{-+})$ (blue circles): $\eta_2(1617)$ and 
$\eta_2(1842)$~\cite{ParticleDataGroup:2024cfk}.
Predicted states (magenta diamonds): $\eta_2(1974)$, $\eta_2(2067)$, $\eta_2(2139)$, and $\eta_2(2199)$ with masses of $1974\pm 13~\mathrm{MeV}$, $2067\pm 
16~\mathrm{MeV}$, $2139\pm 19~\mathrm{MeV}$, and $2199\pm 21~\mathrm{MeV}$. (Note that the uncertainties are so small that many more states can be predicted to high accuracy.)
The solid red curve presents the best-fit result for the 2-states calculation.
The fit parameter $\alpha = 324.6\pm 
11.5~\mathrm{MeV}$.
}
\label{fig:e2} } }
\end{figure}

The logarithmic fit to the BW masses (MeV) of the two known excited states of $\eta_2(2^{-+})$ (blue circles) and four higher excited states (magenta diamonds) projected is shown in Fig.~\ref{fig:e2}.
%--------------------------------------------
%--------------------------------------------

%-------------------------------------------------------------------------------
\subsubsection{$f_4(4^{++})$ Two Known Excited States: Predict Four More}

Two excited states $f_4(4^{++})$ are recorded in the Particle Data
Listings~\cite{ParticleDataGroup:2024cfk}. $f_4(4^{++})$: $I^G~(J^{PC})S = 0^+(4^{++})0$.

\begin{figure}[htb!]
%\vspace{-0.3cm}
\centering
{
    \includegraphics[width=0.5\textwidth,keepaspectratio]{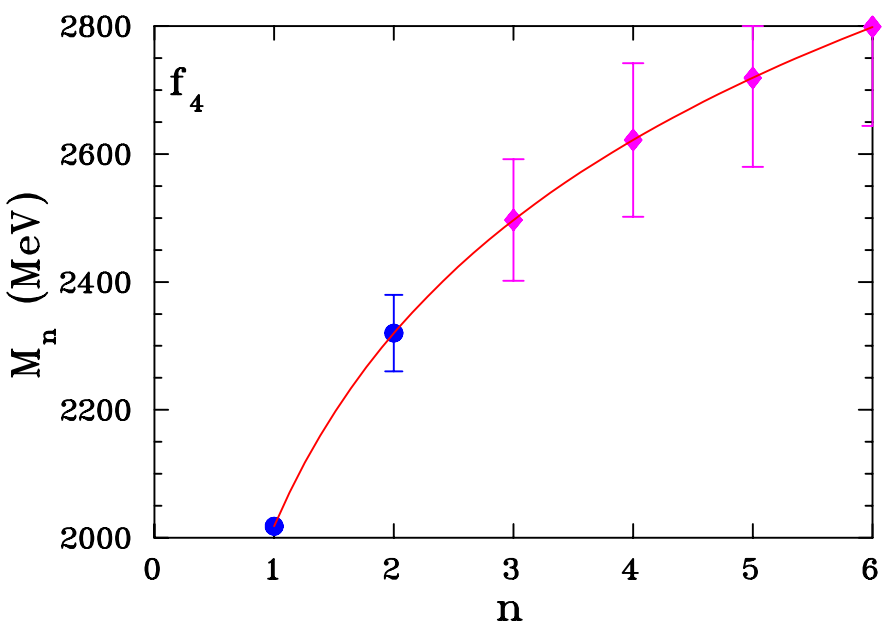} 
}

\centerline{\parbox{0.8\textwidth}{
\caption[] {\protect\small
Data for $f_4(4^{++})$ (blue circles): $f_4(2018)$ and 
$f_4(2320)$~\cite{ParticleDataGroup:2024cfk}.
Predicted states (magenta diamonds): $f_4(2497)$, $f_4(2622)$, $f_4(2719)$, and $f_4(2799)$ with masses of $2497\pm 95~\mathrm{MeV}$, $2622\pm 
120~\mathrm{MeV}$, $2719\pm 139~\mathrm{MeV}$, and $2799\pm 155~\mathrm{MeV}$.
The solid red curve presents the best-fit result for the 2-states calculation.
The fit parameter $\alpha = 435.7\pm 
86.6~\mathrm{MeV}$.
}
\label{fig:f4} } }
\end{figure}

The logarithmic fit to the BW masses (MeV) of the two known excited states of $f_4(4^{++})$ (blue circles) and four higher excited states (magenta diamonds) projected is shown in Fig.~\ref{fig:f4}.
%--------------------------------------------
%--------------------------------------------

%-------------------------------------------------------------------------------
\subsection{Strange Mesons}

%-------------------------------------------------------------------------------
\subsubsection{$K^\ast_2(2^+)$ Two Known Excited States: Predict Four More}

Two excited states $K^\ast_2(2^+)$ are recorded in the Particle Data
Listings~\cite{ParticleDataGroup:2024cfk}. $K^\ast_2(2^+)$: $I(J^P) = 1/2(2^+)$. 

\begin{figure}[htb!]
%\vspace{-0.3cm}
\centering
{
    \includegraphics[width=0.5\textwidth,keepaspectratio]{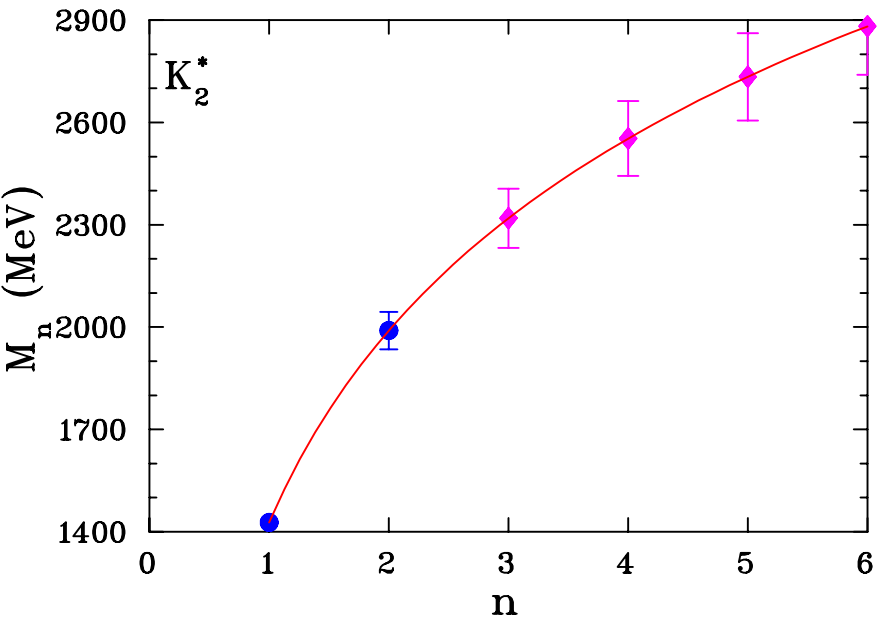} 
}

\centerline{\parbox{0.8\textwidth}{
\caption[] {\protect\small
Data for $K^\ast_2(2^+)$ (blue circles): $K^\ast_2(1427)$ and 
$K^\ast_2(1990)$~\cite{ParticleDataGroup:2024cfk}.
Predicted states (magenta diamonds): $K^\ast_2(2319)$, $K^\ast_2(2553)$, $K^\ast_2(2734)$, and 
$K^\ast_2(2882)$ with masses of $2319\pm 87~\mathrm{MeV}$, $2553\pm 
110~\mathrm{MeV}$, $2734\pm 128~\mathrm{MeV}$, and $2882\pm 142~\mathrm{MeV}$.
The solid red curve presents the best-fit result for the 2-states calculation.
The fit parameter $\alpha = 811.8\pm 
79.3~\mathrm{MeV}$.
}
\label{fig:K2} } }
\end{figure}

The logarithmic fit to the BW masses (MeV) of the two known excited states of $K^\ast_2(2^+)$ (blue circles) and four higher excited states (magenta diamonds) projected is shown in Fig.~\ref{fig:K2}.
%--------------------------------------------
%--------------------------------------------
%\subsubsection{$K^\ast_4(4^+)$ Two Known Excited States: Predict Four More}
%
%Two excited states $K^\ast_4(4^+)$ are recorded in the Particle Data
%Listings~\cite{ParticleDataGroup:2024cfk}. $K^\ast_4(4^+)$: $I(J^P) = 1/2(4^+)$. 
%
%The logarithmic fit to the BW masses (MeV) of the two known excited states of $K^\ast_4(4^+)$ (blue circles) and four higher excited states (magenta diamonds) projected is shown in Fig.~\ref{fig:K4}.
%--------------------------------------------
%\begin{figure}[htb!]
%\vspace{-0.3cm}
%\centering
%{
%    \includegraphics[width=0.5\textwidth,keepaspectratio]{./pig6Y.png} 
%}
%
%\centerline{\parbox{0.8\textwidth}{
%\caption[] {\protect\small
%Data for $K^\ast_4(4^+)$ (blue circles): $K^\ast_4(2048)$ and 
%$K^\ast_4(2490)$~\cite{ParticleDataGroup:2024cfk}.
%Predicted states (magenta diamonds): $K^\ast_4(2748)$, $K^\ast_4(2932)$, $K^\ast_4(3074)$, and 
%$K^\ast_2(3190)$ with masses of $2748\pm 32~\mathrm{MeV}$, $2932\pm 
%40~\mathrm{MeV}$, $3074\pm 46~\mathrm{MeV}$, and $3190\pm 52~\mathrm{MeV}$.
%The solid red curve presents the best-fit result for the 2-states calculation.
%The fit parameter $\alpha = 637.7\pm 
%28.8~\mathrm{MeV}$.
%}
%\label{fig:K4} } }
%\end{figure}
%--------------------------------------------

%-------------------------------------------------------------------------------
\subsection{Charmed Mesons}

%-------------------------------------------------------------------------------
\subsubsection{$D^\ast_1(1^-)$ Two Known Excited States: Predict Four More}

Two excited states are recorded in the Particle Data
Listings~\cite{ParticleDataGroup:2024cfk}. $D^\ast_1(1^-)$: $I(J^P) = 1/2(1^-)$. 

\begin{figure}[htb!]
%\vspace{-0.3cm}
\centering
{
    \includegraphics[width=0.5\textwidth,keepaspectratio]{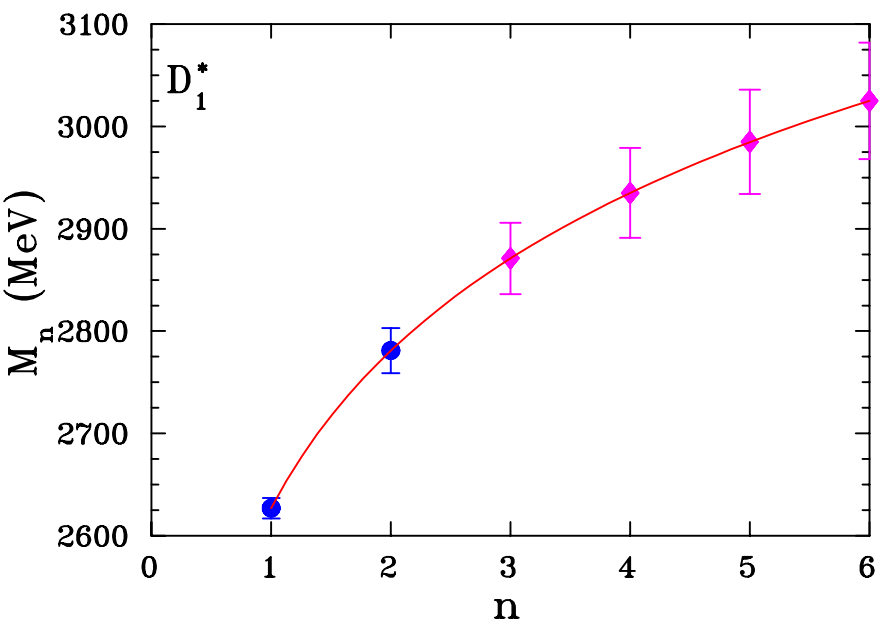} 
}

\centerline{\parbox{0.8\textwidth}{
\caption[] {\protect\small
Data for $D^\ast_1(1^-)$ (blue circles): $D^\ast_1(2627)$ and 
$D^\ast_1(2781)$~\cite{ParticleDataGroup:2024cfk}.
Predicted states (magenta diamonds): $D^\ast_1(2871)$, $D^\ast_1(2935)$, $D^\ast_1(2985)$, and 
$D^\ast_1(3025)$ with masses of $2871\pm 35~\mathrm{MeV}$, $2935\pm 
44~\mathrm{MeV}$, $2985\pm 51~\mathrm{MeV}$, and $3025\pm 57~\mathrm{MeV}$.
The solid red curve presents the best-fit result for the 2-states calculation.
The fit parameter $\alpha = 222.2\pm 
31.7~\mathrm{MeV}$.
}
\label{fig:D11} } }
\end{figure}

The logarithmic fit to the BW masses (MeV) of the two known excited states of $D^\ast_1$ (blue circles) and four higher excited states (red squares) projected is shown in Fig.~\ref{fig:D11}.
%--------------------------------------------
%--------------------------------------------

%-------------------------------------------------------------------------------
\subsection{Charmed Strange Mesons}

%-------------------------------------------------------------------------------
\subsubsection{$D_{s1}(1^+)$ Two Known Excited States: Predict Four More}

Two excited states are recorded in the Particle Data
Listings~\cite{ParticleDataGroup:2024cfk}. $D_{s1}$: $I(J^P) = 0(1^+)$. 

\begin{figure}[htb!]
%\vspace{-0.3cm}
\centering
{
    \includegraphics[width=0.5\textwidth,keepaspectratio]{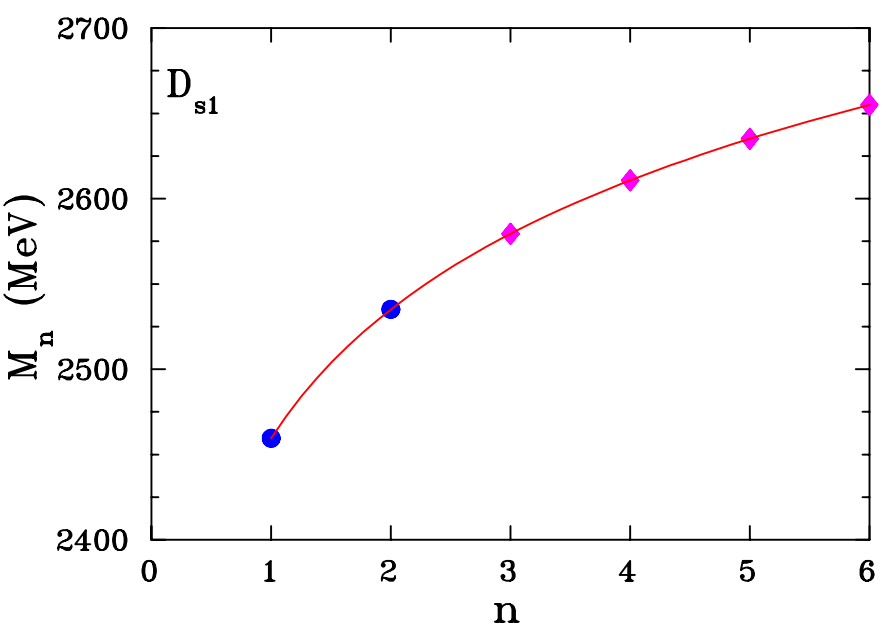} 
}

\centerline{\parbox{0.8\textwidth}{
\caption[] {\protect\small
Data for $D_{s1}(1^+)$ (blue circles): $D_{s1}(2459)$ and 
$D_{s1}(2535)$~\cite{ParticleDataGroup:2024cfk}.
Predicted states (magenta diamonds): $D _{s1}(2579)$, $D_{s1}(2610)$, $D_{s1}(2635)$, and 
$D_{s1}(2655)$ with masses of $2579.34\pm 0.10~\mathrm{MeV}$, $2610.72\pm 0.12~\mathrm{MeV}$, $2635.06\pm 
0.14~\mathrm{MeV}$, and $2655\pm 0.16~\mathrm{MeV}$. (Note that the uncertainties are so small that many more states can be predicted to high accuracy.)
The solid red curve presents the best-fit result for the 2-states calculation.
The fit parameter $\alpha = 109.1\pm 
0.1~\mathrm{MeV}$.
}
\label{fig:D21} } }
\end{figure}

The logarithmic fit to the BW masses (MeV) of the two known excited states of $D _{s1}$ (blue circles) and four higher excited states (red squares) projected is shown in Fig.~\ref{fig:D21}.
%--------------------------------------------
%--------------------------------------------

%-------------------------------------------------------------------------------
\subsubsection{$D^\ast_{s1}(1^-)$ Two Known Excited States: Predict Four More}

Two excited states are recorded in the Particle Data
Listings~\cite{ParticleDataGroup:2024cfk}. $D^\ast_{s1}$: $I(J^P) = 0(1^-)$. 

\begin{figure}[htb!]
%\vspace{-0.3cm}
\centering
{
    \includegraphics[width=0.5\textwidth,keepaspectratio]{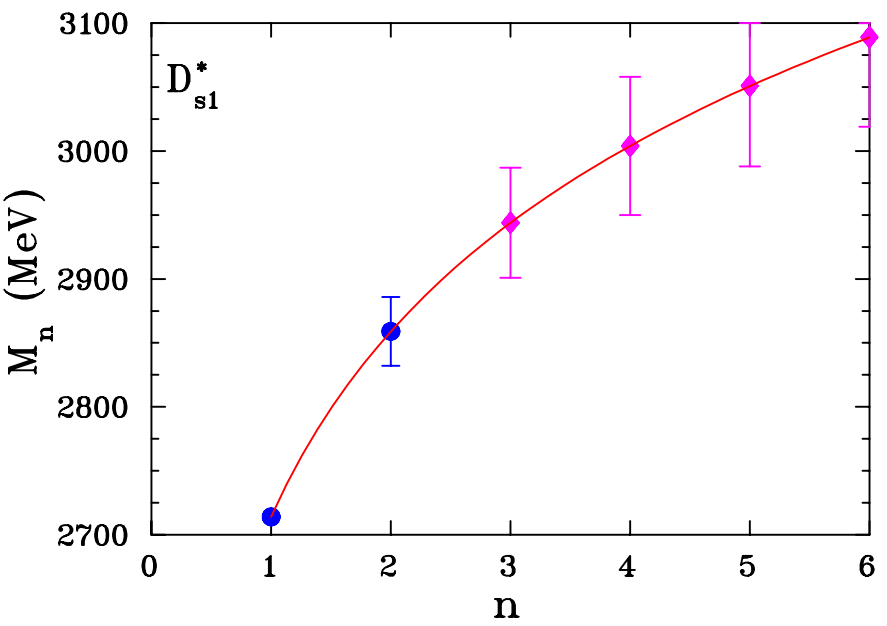} 
}

\centerline{\parbox{0.8\textwidth}{
\caption[] {\protect\small
Data for $D _{s1}(1^-)$ (blue circles): $D^\ast_{s1}(2714)$ and 
$D^\ast_{s1}(2859)$~\cite{ParticleDataGroup:2024cfk}.
Predicted states (magenta diamonds): $D^\ast_{s1}(2944)$, $D^\ast_{s1}(3004)$, $D^\ast_{s1}(3051)$, and 
$D^\ast_{s1}(3089)$ with masses of $2944\pm 43~\mathrm{MeV}$, $3004\pm 54~\mathrm{MeV}$, $3051\pm 
63~\mathrm{MeV}$, and $3089\pm 70~\mathrm{MeV}$.
The solid red curve presents the best-fit result for the 2-states calculation.
The fit parameter $\alpha = 209.2\pm 
38.9~\mathrm{MeV}$.
}
\label{fig:D31} } }
\end{figure}

The logarithmic fit to the BW masses (MeV) of the two known excited states of $D^\ast_{s1}$ (blue circles) and four higher excited states (red squares) projected is shown in Fig.~\ref{fig:D31}.
%--------------------------------------------
%--------------------------------------------

%-------------------------------------------------------------------------------
\subsection{Bottom Charmed Mesons}

%-------------------------------------------------------------------------------
\subsubsection{$B_c(0^-)$ Two Known Excited States: Predict Four More}

Two excited states are recorded in the Particle Data
Listings~\cite{ParticleDataGroup:2024cfk}. $B_c(0^-)$: $I(J^P) = 0(0^-)$. 

\begin{figure}[htb!]
%\vspace{-0.3cm}
\centering
{
    \includegraphics[width=0.5\textwidth,keepaspectratio]{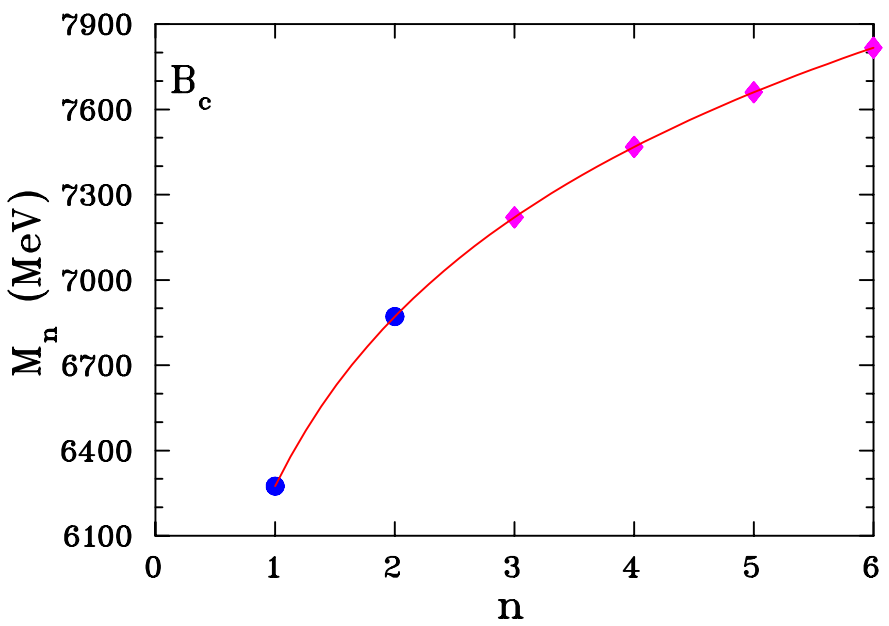} 
}

\centerline{\parbox{0.8\textwidth}{
\caption[] {\protect\small
Data for $B_c(0^-)$ (blue circles): $B_c(6274)$ and 
$B_c(6871)$~\cite{ParticleDataGroup:2024cfk}.
Predicted states (magenta diamonds): $B_c(7220)$, $B_c(7468)$, $B_c(7660)$, and $B_c(7817)$ with masses of $7220.3\pm 1.6~\mathrm{MeV}$, $7467.9\pm 2.0~\mathrm{MeV}$, $7660.0\pm 
2.3~\mathrm{MeV}$, and $7817.0\pm 2.6~\mathrm{MeV}$. (Note that the uncertainties are so small that many more states can be predicted to high accuracy.)
The solid red curve presents the best-fit result for the 2-states calculation.
The fit parameter $\alpha = 860.9\pm 
1.4~\mathrm{MeV}$.
}
\label{fig:B1} } }
\end{figure}

The logarithmic fit to the BW masses (MeV) of the two known excited states of $B _c(0^-)$ (blue circles) and four higher excited states (red squares) projected is shown in Fig.~\ref{fig:B1}.
%--------------------------------------------
%--------------------------------------------

%-------------------------------------------------------------------------------
\subsection{$c\bar{c}$ Mesons}

%-------------------------------------------------------------------------------
\subsubsection{$\eta_c(0^{-+})$ Two Known Excited States: Predict Four More}

Two excited states are recorded in the Particle Data
Listings~\cite{ParticleDataGroup:2024cfk}. $\eta_c(0^{-+})$: $I^G(J^{PC}) = 0^+(0^{-+})$. 

\begin{figure}[htb!]
%\vspace{-0.3cm}
\centering
{
    \includegraphics[width=0.5\textwidth,keepaspectratio]{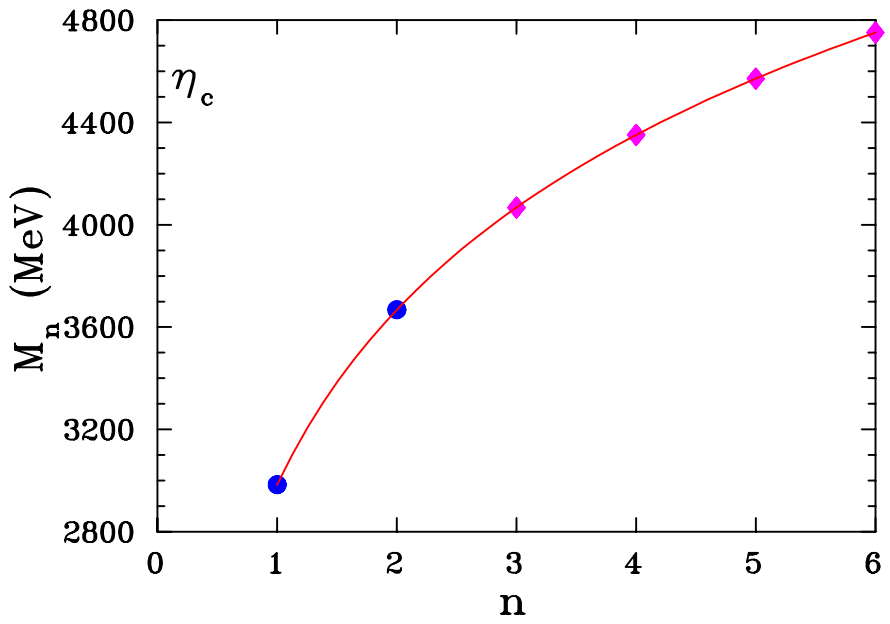} 
}

\centerline{\parbox{0.8\textwidth}{
\caption[] {\protect\small
Data for $\eta_c(0^{-+})$ (blue circles): $\eta_c(2984)$ and 
$\eta_c(3668)$~\cite{ParticleDataGroup:2024cfk}.
Predicted states (magenta diamonds): $\eta_c(4068)$, $\eta_c(4351)$, $\eta_c(4571)$, and 
$\eta_c(4751)$ with masses of $4067.5\pm 1.4~\mathrm{MeV}$, $4351.2\pm 1.8~\mathrm{MeV}$, $4571.2\pm 
2.1~\mathrm{MeV}$, and $4751.0\pm 2.3~\mathrm{MeV}$. (Note
that the uncertainties are so small that many more states can be predicted to high accuracy.)
The solid red curve presents the best-fit result for the 2-states calculation.
The fit parameter $\alpha = 986.2\pm 
1.3~\mathrm{MeV}$.
}
\label{fig:C1} } }
\end{figure}

The logarithmic fit to the BW masses (MeV) of the two known excited states of $\eta_c(0^{-+})$ (blue circles) and four higher excited states (red squares) projected is shown in Fig.~\ref{fig:C1}.
%--------------------------------------------
%--------------------------------------------

%-------------------------------------------------------------------------------
\subsubsection{$\chi_{c2}(2^{++})$ Two Known Excited States: Predict Four More}

Two excited states are recorded in the Particle Data
Listings~\cite{ParticleDataGroup:2024cfk}. $\chi_{c2}(2^{++})$: $I^G(J^{PC}) = 0^+(2^{++})$. (Analog to the $q\bar{q}\;f_2$ meson.)

\begin{figure}[htb!]
%\vspace{-0.3cm}
\centering
{
    \includegraphics[width=0.5\textwidth,keepaspectratio]{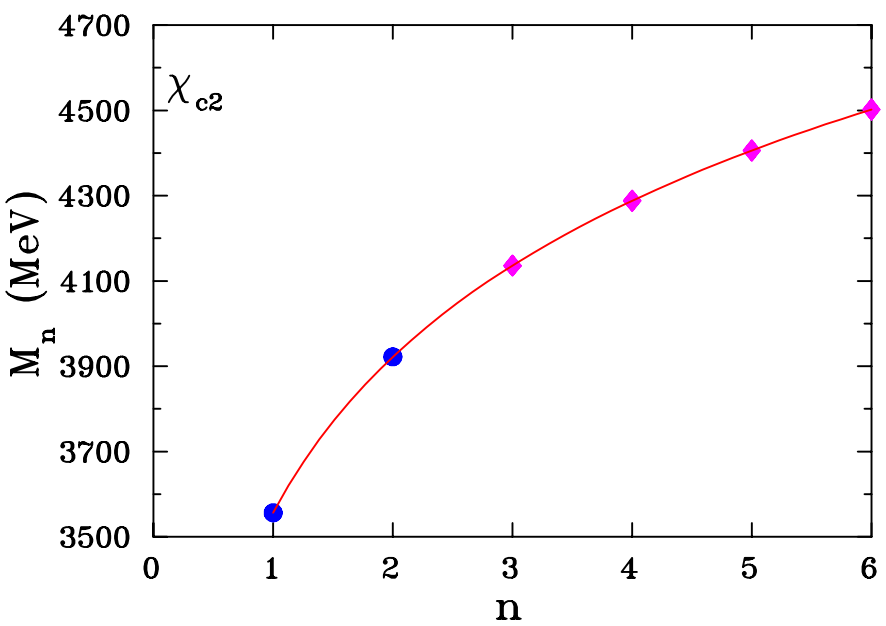} 
}

\centerline{\parbox{0.8\textwidth}{
\caption[] {\protect\small
Data for $\chi_{c2}(2^{++})$ (blue circles): $\chi_{c2}(3556)$ and 
$\chi_{c2}(3922)$~\cite{ParticleDataGroup:2024cfk}.
Predicted states (magenta diamonds): $\chi_{c2}(4136)$, $\chi_{c2}(4288)$, $\chi_{c2}(4406)$, and 
$\chi_{c2}(4502)$ with masses of $4136.1\pm 2.8~\mathrm{MeV}$, $4288.0\pm 3.6~\mathrm{MeV}$, $4405.8\pm 
4.2~\mathrm{MeV}$, and $4502.1\pm 4.6~\mathrm{MeV}$. (Note that the uncertainties are so small that many more states can be predicted to high accuracy.)
The solid red curve presents the best-fit result for the 2-states calculation.
The fit parameter $\alpha = 527.9\pm 
2.6~\mathrm{MeV}$.
}
\label{fig:C2} } }
\end{figure}

The logarithmic fit to the BW masses (MeV) of the two known excited states of $\chi_{c2}(2^{++})$ (blue circles) and four higher excited states (red squares) projected is shown in Fig.~\ref{fig:C2}.
%--------------------------------------------
%--------------------------------------------

%-------------------------------------------------------------------------------
\subsubsection{$T_{c\bar{c}\bar{s}1}(1^+)$ Two Known Excited States: Predict Four More}

Two excited states are recorded in the Particle Data
Listings~\cite{ParticleDataGroup:2024cfk}. $T_{c\bar{c}\bar{s}1}(1^+)$: $I(J^{PC}) = 1/2(1^+)$. 

\begin{figure}[htb!]
%\vspace{-0.3cm}
\centering
{
    \includegraphics[width=0.5\textwidth,keepaspectratio]{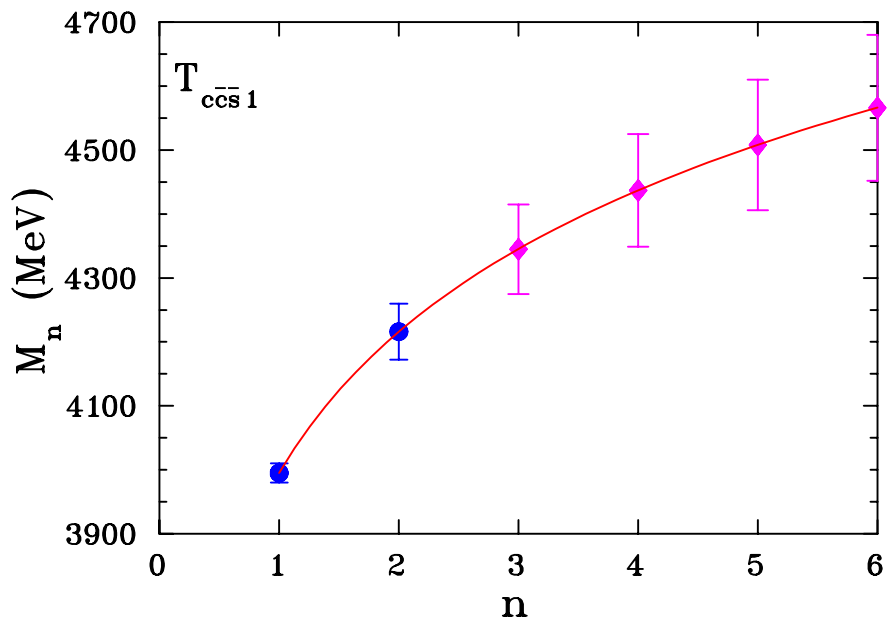} 
}

\centerline{\parbox{0.8\textwidth}{
\caption[] {\protect\small
Data for $T_{c\bar{c}\bar{s}1}(1^+)$ (blue circles): $T_{c\bar{c}\bar{s}1}(3995)$ and 
$T_{c\bar{c}\bar{s}1}(4216)$~\cite{ParticleDataGroup:2024cfk}.
Predicted states (magenta diamonds): $T_{c\bar{c}\bar{s}1}(4345)$, $T_{c\bar{c}\bar{s}1}(4437)$, $T_{c\bar{c}\bar{s}1}(4508)$, and 
$T_{c\bar{c}\bar{s}1}(4566)$ with masses of $4345\pm 70~\mathrm{MeV}$, $4437\pm 88~\mathrm{MeV}$, $4508\pm 102~\mathrm{MeV}$, and $4566\pm 114~\mathrm{MeV}$. The solid red curve presents the best-fit result for the 2-states calculation. The fit parameter $\alpha = 318.8\pm 63.5~\mathrm{MeV}$.
}
\label{fig:T2} } }
\end{figure}

The logarithmic fit to the BW masses (MeV) of the two known excited states of $T_{c\bar{c}\bar{s}1}(1^+)$ (blue circles) and four higher excited states (red squares) projected is shown in Fig.~\ref{fig:T2}.
%--------------------------------------------
%--------------------------------------------

%-------------------------------------------------------------------------------
\subsection{$b\bar{b}$ Mesons}

%-------------------------------------------------------------------------------
\subsubsection{$\eta_b(0^{-+})$ Two Known Excited States: Predict Four More}

Two excited states $\eta_b(0^{-+})$ are recorded in the Particle Data
Listings~\cite{ParticleDataGroup:2024cfk}. $\eta_b(0^{-+})$: $I^G(J^{PC}) = 0^+(0^{-+})$. 

\begin{figure}[htb!]
%\vspace{-0.3cm}
\centering
{
    \includegraphics[width=0.5\textwidth,keepaspectratio]{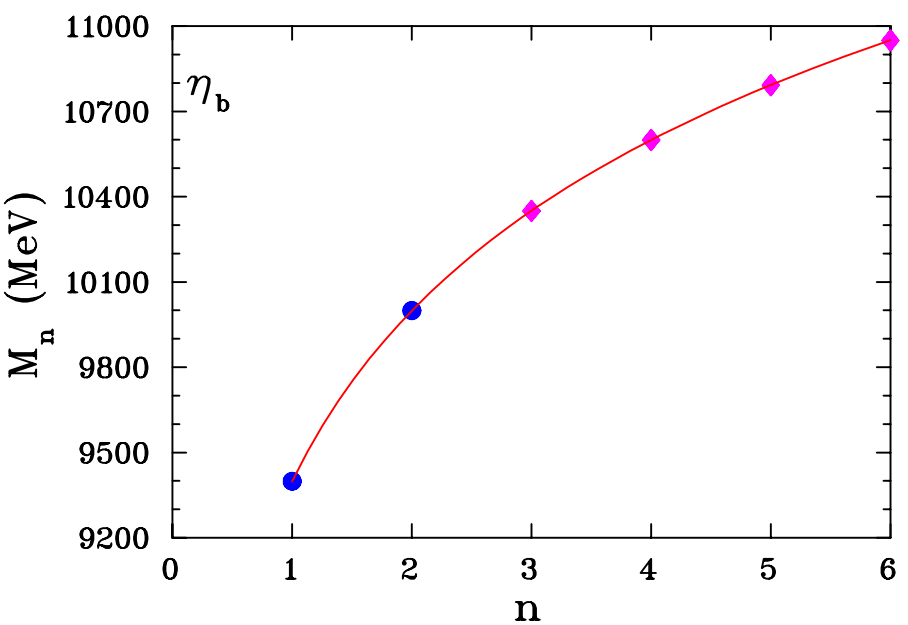} 
}

\centerline{\parbox{0.8\textwidth}{
\caption[] {\protect\small
Data for $\eta_b(0^{-+})$ (blue circles): $\eta_b(9399)$ and 
$\eta_b(9999)$~\cite{ParticleDataGroup:2024cfk}.
Predicted states (magenta diamonds): $\eta_b(10350)$, $\eta_b(10599)$, $\eta_b(10792)$, and 
$\eta_b(10950)$ with masses of $10350\pm 7~\mathrm{MeV}$, $10599\pm 8~\mathrm{MeV}$, $10792\pm 
10~\mathrm{MeV}$, and $10950\pm 11~\mathrm{MeV}$. (Note that the uncertainties are so small that many more states can be predicted to high accuracy.)
The solid red curve presents the best-fit result for the 2-states calculation.
The fit parameter $\alpha = 866.0\pm 
6.1~\mathrm{MeV}$.
}
\label{fig:29} } }
\end{figure}

The logarithmic fit to the BW masses (MeV) of the two known excited states of $\eta_b(0^{-+})$ (blue circles) and four higher excited states (red squares) projected is shown in Fig.~\ref{fig:29}.
%--------------------------------------------
%--------------------------------------------

%-------------------------------------------------------------------------------
\subsubsection{$h_b(1^{+-})$ Two Known Excited States: Predict Four More}

Two excited states are recorded in the Particle Data
Listings~\cite{ParticleDataGroup:2024cfk}. $h_b(1^{+-})$: $(I,J^P) = 0^-(1^{+-})$. 

\begin{figure}[htb!]
%\vspace{-0.3cm}
\centering
{
    \includegraphics[width=0.5\textwidth,keepaspectratio]{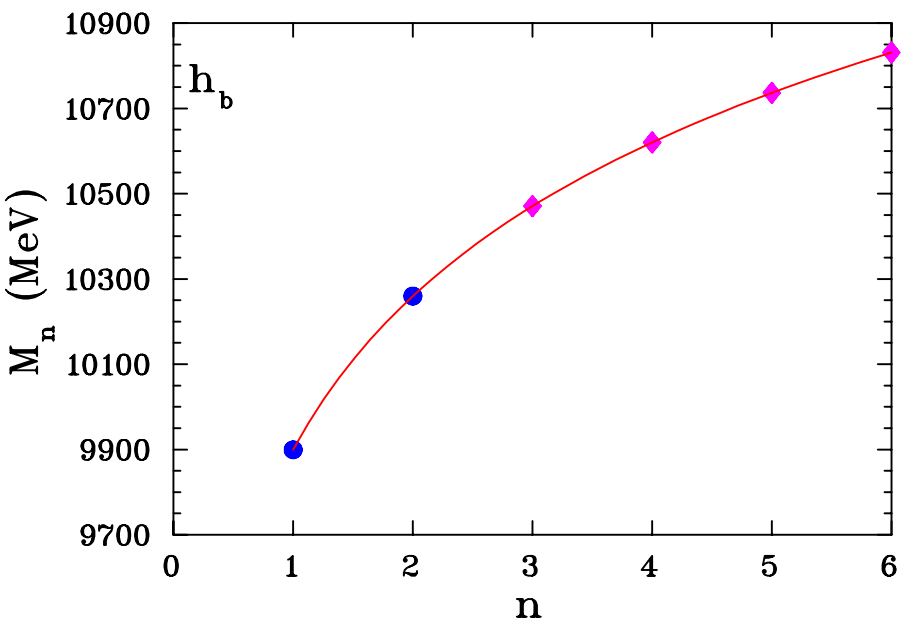} 
}

\centerline{\parbox{0.8\textwidth}{
\caption[] {\protect\small
Data for $h_b(1^{+-})$ (blue circles): $h_b(9899)$ and 
$h_b(10260)$~\cite{ParticleDataGroup:2024cfk}.
Predicted states (magenta diamonds): $h_b(10471)$, $h_b(10620)$, $h_b(10736)$, and 
$h_b(10831)$ with masses of $10470.7\pm 1.9~\mathrm{MeV}$, $10620.3\pm 2.4~\mathrm{MeV}$, $10736.3\pm 
2.8~\mathrm{MeV}$, and $10831.2\pm 3.1~\mathrm{MeV}$. (Note that the uncertainties are so small that many more states can be predicted to high accuracy.)
The solid red curve presents the best-fit result for the 2-states calculation.
The fit parameter $\alpha = 520.1\pm 
1.7~\mathrm{MeV}$.
}
\label{fig:E5} } }
\end{figure}

The logarithmic fit to the BW masses (MeV) of the two known excited states of $h_b(1^{+-})$ (blue circles) and four higher excited states (red squares) projected is shown in Fig.~\ref{fig:E5}.
%--------------------------------------------
%--------------------------------------------

%-------------------------------------------------------------------------------
\subsubsection{$\chi_{b0}(0^{++})$ Two Known Excited States: Predict Four More}

Two excited states are recorded in the Particle Data
Listings~\cite{ParticleDataGroup:2024cfk}. $\chi_{b0}(0^{++})$: $I^G(J^{PC}) = 0^+(0^{++})$. (Analog to the $q\bar{q}\;f_0$ meson.)

\begin{figure}[htb!]
%\vspace{-0.3cm}
\centering
{
    \includegraphics[width=0.5\textwidth,keepaspectratio]{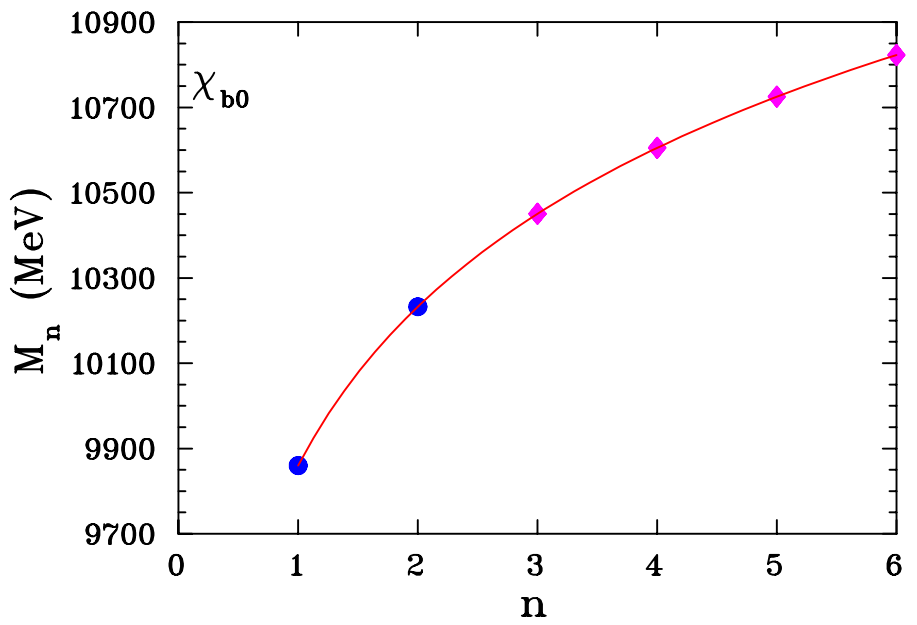} 
}

\centerline{\parbox{0.8\textwidth}{
\caption[] {\protect\small
Data for $\chi_{b0}(0^{++})$ (blue circles): $\chi_{b0}(9860)$ and 
$\chi_{b0}(10232)$~\cite{ParticleDataGroup:2024cfk}.
Predicted states (magenta diamonds): $\chi_{b0}(10450)$, $\chi_{b0}(10605)$, $\chi_{b0}(10725)$, and 
$\chi_{b0}(10823)$ with masses of $10450.4\pm 1.0~\mathrm{MeV}$, $10605.1\pm 1.3\mathrm{MeV}$, $10725.0\pm 
1.5~\mathrm{MeV}$, and $10823.0\pm 1.6~\mathrm{MeV}$. (Note that the uncertainties are so small that many more states can be predicted to high accuracy.)
The solid red curve presents the best-fit result for the 2-states calculation.
The fit parameter $\alpha = 537.5\pm 
0.9~\mathrm{MeV}$.
}
\label{fig:C6} } }
\end{figure}

The logarithmic fit to the BW masses (MeV) of the two known excited states of $\chi _{b0}(0^{++})$ (blue circles) and four higher excited states (red squares) projected is shown in Fig.~\ref{fig:C6}.
%--------------------------------------------
%--------------------------------------------

%-------------------------------------------------------------------------------
%\subsection{Other Mesons}

%-------------------------------------------------------------------------------
\subsubsection{$T_{b\bar{b}1}(1^{+-})$ Two Known Excited States: Predict Four More}

Two excited states are recorded in the Particle Data
Listings~\cite{ParticleDataGroup:2024cfk}. $T_{b\bar{b}1}(1^{+-})$: $I^G(J^{PC}) = 1^+(1^{+-})$. (Analog to the $q\bar{q}\;b_1$ meson.)

\begin{figure}[htb!]
%\vspace{-0.3cm}
\centering
{
    \includegraphics[width=0.5\textwidth,keepaspectratio]{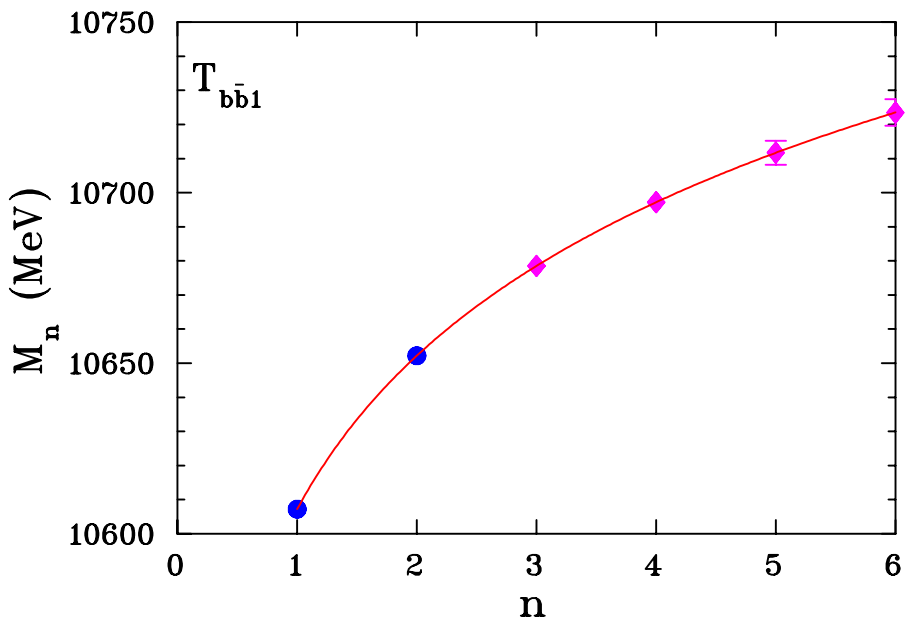} 
}

\centerline{\parbox{0.8\textwidth}{
\caption[] {\protect\small
Data for $T_{b\bar{b}1}(1^{+-})$ (blue circles): $T_{b\bar{b}1}(10607)$ and 
$T_{b\bar{b}1}(10652)$~\cite{ParticleDataGroup:2024cfk}.
Predicted states (magenta diamonds): $T_{b\bar{b}1}(10679)$, $T_{b\bar{b}1}(10697)$, $T_{b\bar{b}1}(10712)$, and 
$T_{b\bar{b}1}(10724)$ with masses of $10678.5\pm 2.4~\mathrm{MeV}$, $10697.2\pm 3.0~\mathrm{MeV}$, $10711.7\pm 
3.5~\mathrm{MeV}$, and $10723.5\pm 3.9~\mathrm{MeV}$. (Note that the uncertainties are so small that many more states can be predicted to high accuracy.)
The solid red curve presents the best-fit result for the 2-states calculation.
The fit parameter $\alpha = 64.9\pm 
2.2~\mathrm{MeV}$.
}
\label{fig:T1} } }
\end{figure}

The logarithmic fit to the BW masses (MeV) of the two known excited states of $T_{b\bar{b}1}(1^{+-})$ (blue circles) and four higher excited states (red squares) projected is shown in Fig.~\ref{fig:T1}.
%--------------------------------------------
%--------------------------------------------

%-------------------------------------------------------------------------------
\subsection{Cumulative Meson Excited States}
\raggedright

\begin{figure}[htb!]
%\vspace{-0.3cm}
\centering
{
   \includegraphics[width=0.45\textwidth,keepaspectratio]{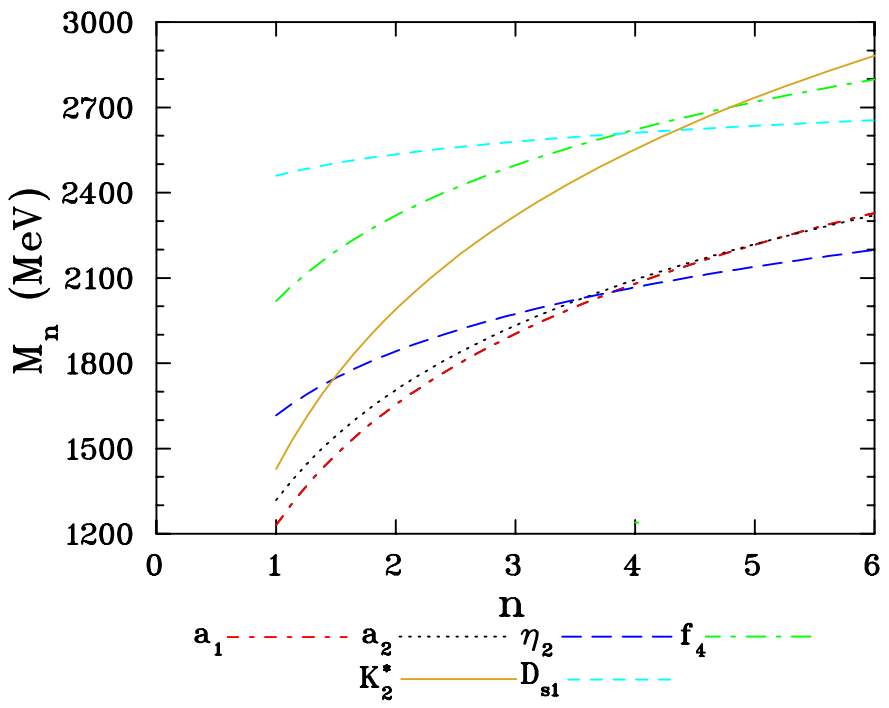}~~~ 
   \includegraphics[width=0.44\textwidth,keepaspectratio]{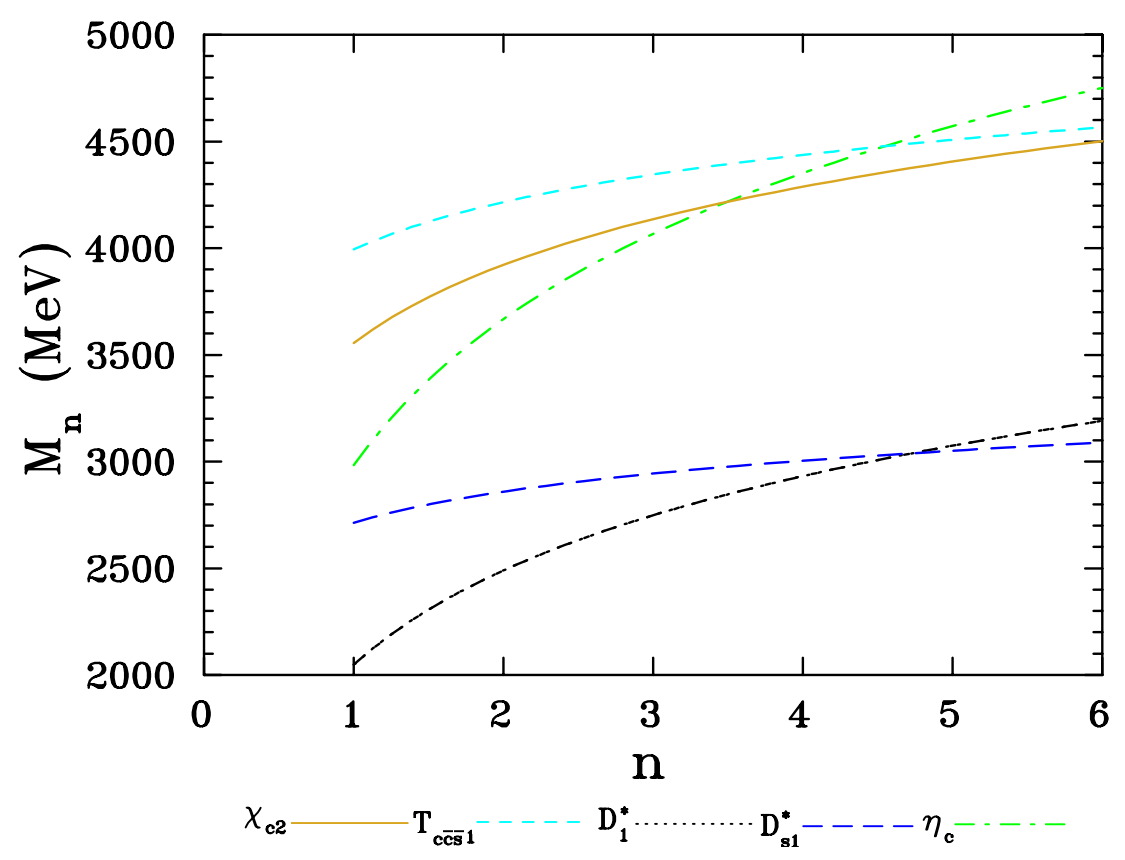}
   \includegraphics[width=0.43\textwidth,keepaspectratio]{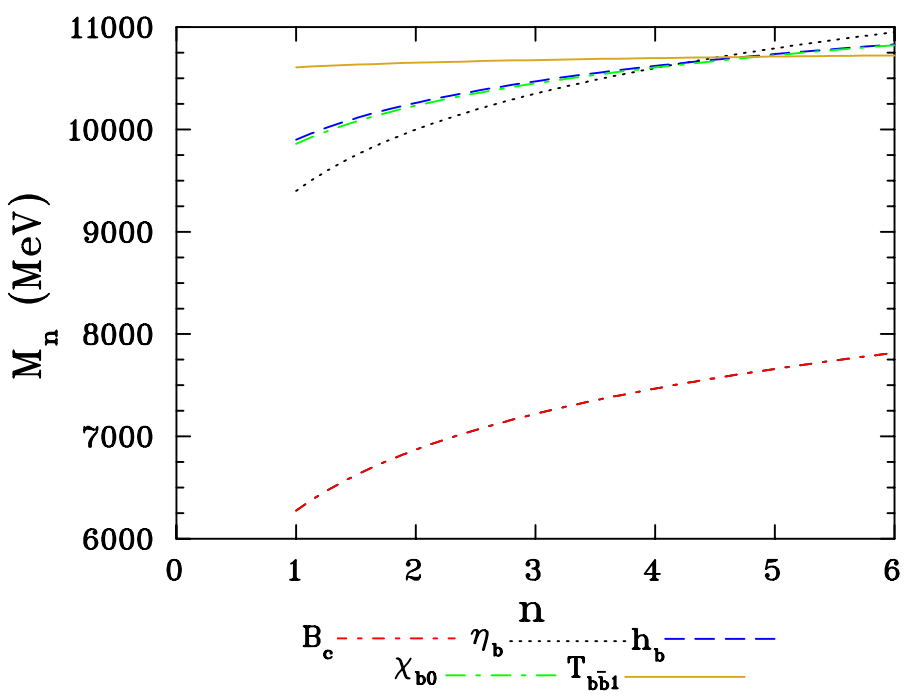}
}

\centerline{\parbox{0.8\textwidth}{
\caption[] {\protect\small
Logarithmic curves for sixteen meson data sets that only have two known excited states.
\underline{Top Left}: $a_1(1^{++})$, $a_2(2^{++})$, $\eta_2(2^{++})$, 
$f_4(4^{++})$, $K^\ast_2(2^+)$, and $D_{s1}(1^+)$.
\underline{Top Right}: 
%$K^\ast_4(4^+)$, 
$D^\ast_1(1^-)$, $D^\ast_{s1}(1^-)$,
$\eta_c(0^{-+})$, $\chi_{c2}(2^{++})$, and $T_{c\bar{c}\bar{s}1}(1^+)$.
\underline{Bottom}: $B_c(0^-)$, $\eta_b(0^{-+})$, $h_b(1^{+-})$, $\chi_{b0}(0^{++})$, and $T_{b\bar{b}1}(1^{+-})$.
}
\label{fig:peg} } }
\end{figure}

The cumulative fit curves of sixteen mesons with sets of equal-quantum mass with only two excited states per set are shown in Fig.~\ref{fig:peg}. The logarithmic slope, $\alpha$, usually decreases as the mass of the ground state increases.

%--------------------------------------------
%--------------------------------------------

\begin{figure}[htb!]
%\vspace{-0.3cm}
\centering
{
    \includegraphics[width=0.5\textwidth,keepaspectratio]{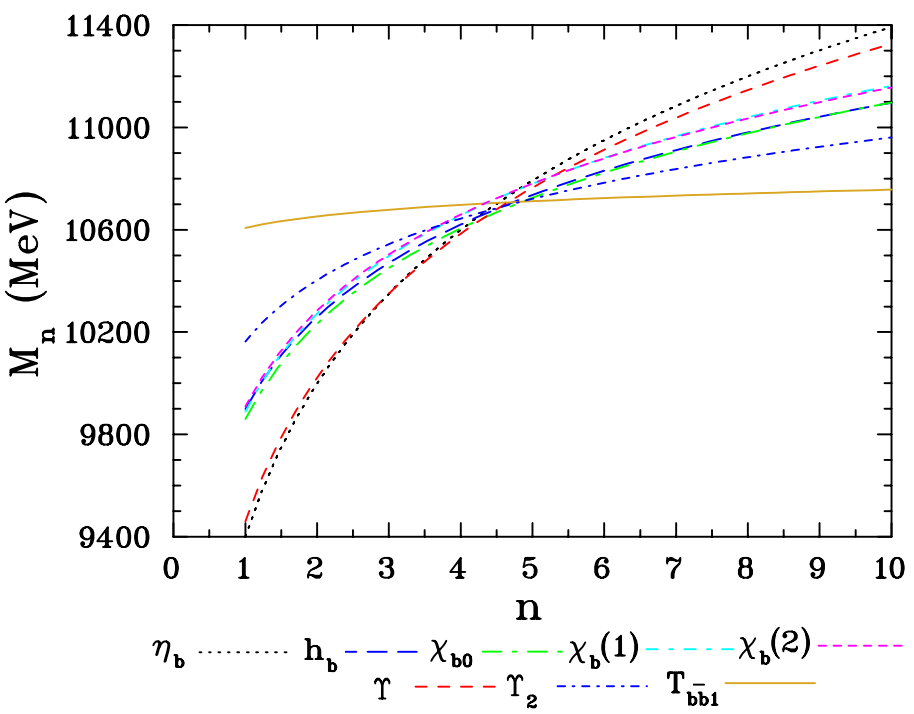} 
}

\centerline{\parbox{0.8\textwidth}{
\caption[] {\protect\small
Eight $b\bar{b}$ meson sets from Part~I~\cite{Roper:2024ovj} and current Part~II. The near crossover of all curves at one point occurs because of the near linear relationship between the $\alpha$ parameter and the ground-state mass shown in Fig.~35. If the linear relationship were perfect for all $b\bar{b}$ excited-states sets, the crossover would occur at $n_c=5.10$ and $M_{n_c}= 10768~\mathrm{MeV}$, as determined by a fit to a perfect crossover. (Fit parameters: $\gamma_c=-0.6136~\mathrm{MeV}$ and $\mu_c = 6607.5~\mathrm{MeV}$. See Eq.~(\ref{eq:eqbbA}) for $\alpha_c(M_1)$ to calculate the best-fit state masses.)}
\label{fig:bbarb} } }
\end{figure}

Combine the set of heavy quark $b\bar{b}$ mesons (Fig.~\ref{fig:bbarb}). They nearly have a common crossing point corresponding to $n_c \approx 4.82$ or $r_c \approx 0.13~\mathrm{fm}$.  The crossover mass is $M_c \approx 10708~\mathrm{MeV}$.
%--------------------------------------------
%--------------------------------------------

The seven $b\bar{b}$ excited-states sets from Parts~I~\cite{Roper:2024ovj} and II are listed in Table~\ref{tbl:tabbb}. 
%-------------------------------------------------
\begin{table}[htb!]

\centering \protect\caption{{List of $b\bar{b}$ excited-states sets, maximum $n = 3-7$, in Parts~I~\cite{Roper:2024ovj} and, maximum $n = 2$, Part~II (current).}}

\vspace{2mm}
{%
\begin{tabular}{|c|c|c|c|}
\hline
Meson              &  $\alpha$       &  $\beta$       & \# of  \tabularnewline
                   & (MeV)           & (MeV)          & States \tabularnewline
\hline
$T_{b\bar{b}1}(1^{+-})$& 64.9$\pm$2.2& 10607          & 2~states \tabularnewline
$h_b(1^{+-})$      & 520.1$\pm$1.7   & 9899           & 2~states \tabularnewline
$\chi_{b0}(0^{++})$& 537.5$\pm$0.9   & 9860           & 2~states \tabularnewline
$\chi_{b2}(2^{++})$& 540.3$\pm$19.4  & 9908.8$\pm$11.2& 3~states \tabularnewline
$\chi_{b1}(1^{++})$& 553.2$\pm$26.0  & 9889.4$\pm$15.6& 3~states \tabularnewline
$\Upsilon(1^{-~-})$& 810.6$\pm$3.4   & 9460.5$\pm$1.0 & 7~states \tabularnewline
$\eta_b(0^{-+})$   & 866.0$\pm$6.1   & 9399           & 2~states \tabularnewline
\hline
\end{tabular}} \label{tbl:tabbb}
\end{table}
%--------------------------------------------
The linear trend of $b\bar{b}$ sets' $\alpha$ \textit{vs.} mass linear trend in Fig.~35, because of its strong linearity, can be used to accurately estimate the parameter $\alpha$ for quantum states $b\bar{b}$ for which only the ground-state mass is known, to be used to calculate the higher masses of the set. The $\alpha(M_1)$ equation for the $b\bar{b}$ sets' linear trend is
\begin{equation}
    \alpha(M_{1}) = \gamma M_1 + \mu \equiv [(-0.6359\pm 0.0028) M_{1} + (6809.3\pm 27.4)]~\mathrm{MeV} \>.
\label{eq:eqbbA}
\end{equation}
(The input uncertainties are $\delta M_1,~\delta\gamma=0.0028)$ and $\delta\mu=27.4.$)

Then, Eq.~(\ref{eq:eq1})  becomes
\begin{equation}
    M_n = (\gamma M_1 + \mu) Ln(n) + M_1 = 
    [\gamma~Ln(n)~+~1]M_1 + \mu~Ln(n)\>,
\label{eq:eqbb}
\end{equation}
an accurate one-parameter ($M_1$) equation for $b\bar{b}$ excited-states sets. (The output uncertainty is $\delta M_n$, calculated using Eq.~(\ref{eq:eq4}) and $\delta \alpha = \sqrt{ \gamma^2(\delta M_1)^2 + (M_1)^2(\delta\gamma)^2 + (\delta \mu)^2}$ and $\delta \beta = \delta M_1$.)

%---------------------------------------------------------
\begin{figure}[htb!]
\vspace{-0.3cm}
\centering
{
    \includegraphics[width=0.6\textwidth,keepaspectratio]{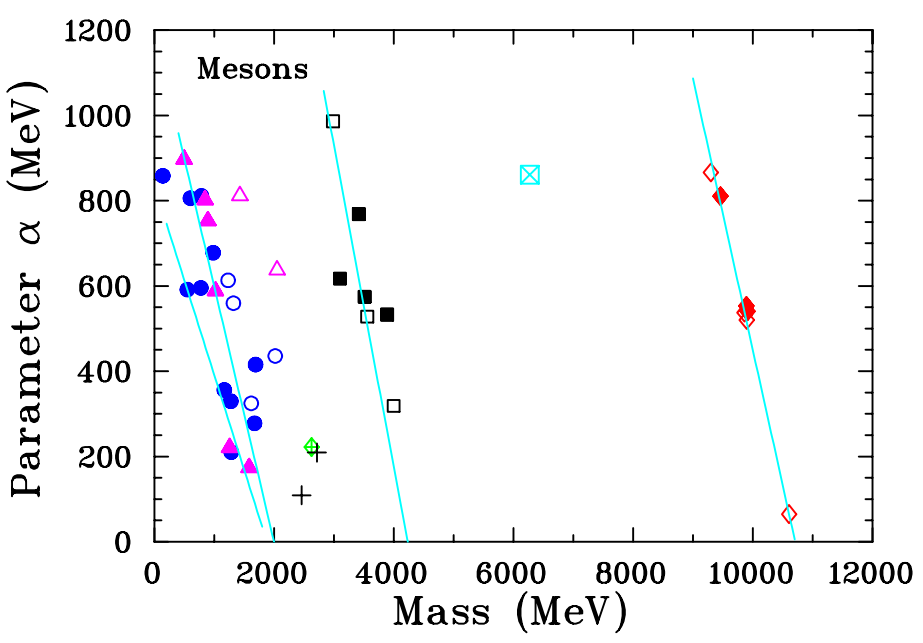}
}
\centerline{\parbox{0.8\textwidth}{
\caption[] {\protect\small
Fit parameter $\alpha$ versus $\beta$ = mass of the ground-state meson. This illustrates an approximate inverse relationship of the fit parameter $\alpha$ with the ground-state mass. There are 40 meson sets in this graph, 16 from this study and 24 from SR. Filled (open) symbols correspond to more than 2-states in sets from SK (2-states sets from this study). Light unflavored (strange) mesons shown by blue circles (magenta triangles),
$c\bar{c}$ ($b\bar{b}$) mesons shown by black squares (red diamonds),
charmed (charmed strange) mesons shown by green diamond with cross (black cross), and
bottom charmed meson shown by cyan square with cross. The cyan thin lines are for the $ud$, $s\bar{s}$, $c\bar{c}$, and $b\bar{b}$ sets' linear trends, respectively. 
Eq.~(\ref{eq:eqbbA}) is the equation for the $b\bar{b}$ line on the right. }}
\label{fig:avb} } 
\end{figure}
%----------------------------------------------------------

In a sense, Eq.~(\ref{eq:eqbb}) is a ``zero-parameter'' equation since $M_1$ is the ground-state ``definition'' of a $b\bar{b}$ set. In other words, once the ground-state mass is known, all higher-state masses are known.

State masses calculated by Eq.~(\ref{eq:eq1}) and Eq.~(\ref{eq:eqbb}) for the seven heavy quark $b\bar{b}$ sets listed in Table~\ref{tbl:tabbb} differ at most by $0.79\%$; most are much smaller.

Note that $\alpha=0$ in Eq.~(\ref{eq:eqbbA}) yields $M_1=10708$ MeV, the apparent maximum possible value for a ground-state mass, if Eq.~(\ref{eq:eqbbA}) is valid.

\begin{figure}[htb!]
%\vspace{-0.3cm}
\centering
{
    \includegraphics[width=0.45\textwidth,keepaspectratio]{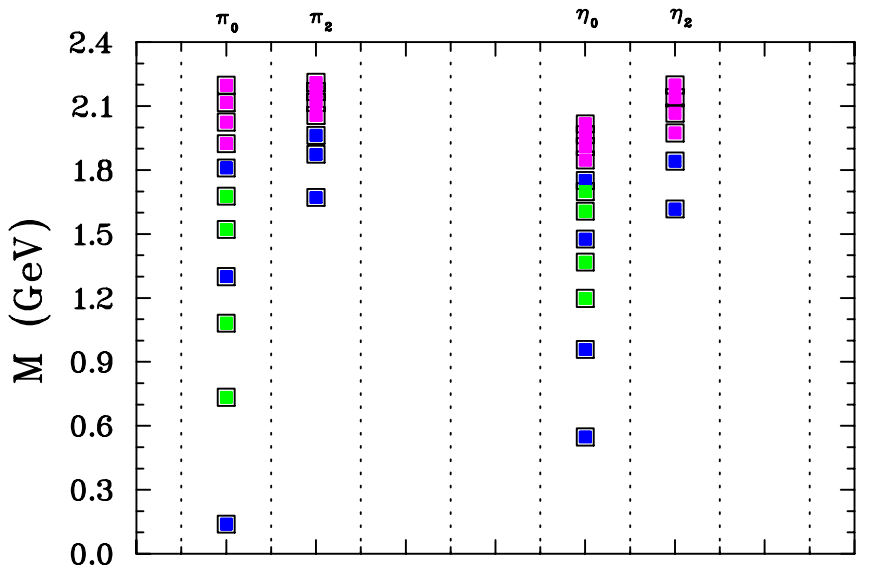}~~~ 
    \includegraphics[width=0.45\textwidth,keepaspectratio]{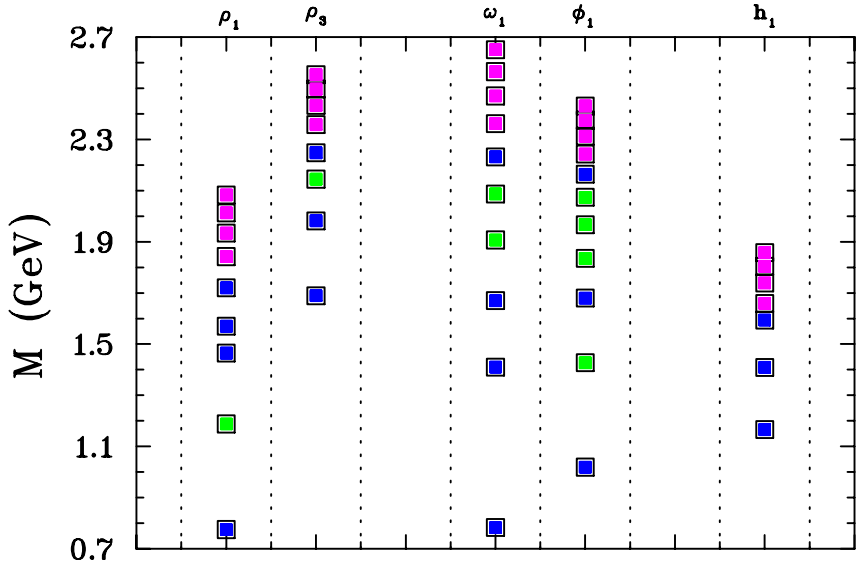}
    \includegraphics[width=0.45\textwidth,keepaspectratio]{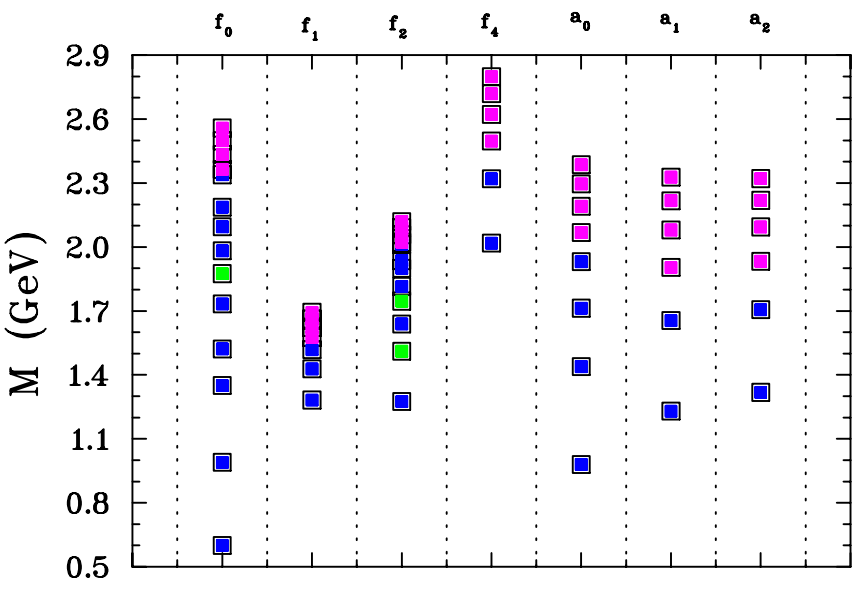}~~~ 
    \includegraphics[width=0.45\textwidth,keepaspectratio]{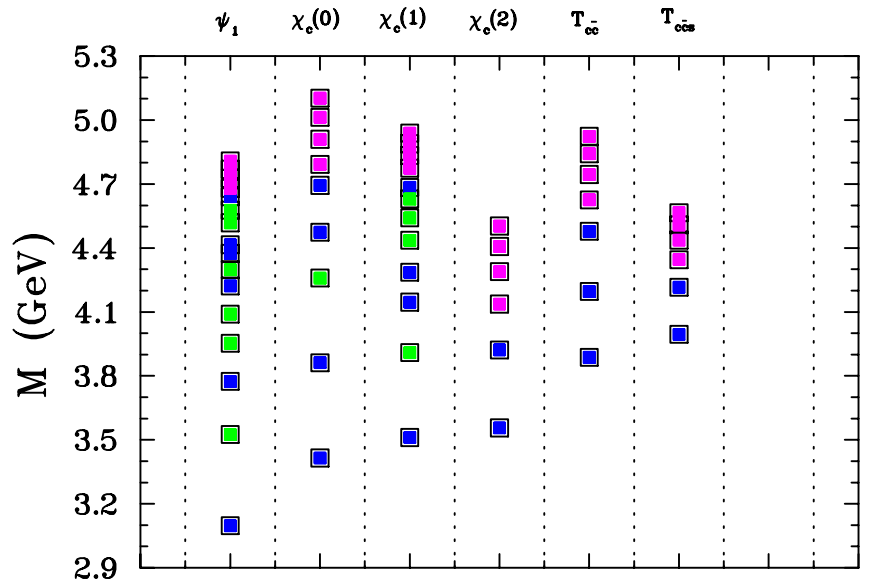} 
    \includegraphics[width=0.45\textwidth,keepaspectratio]{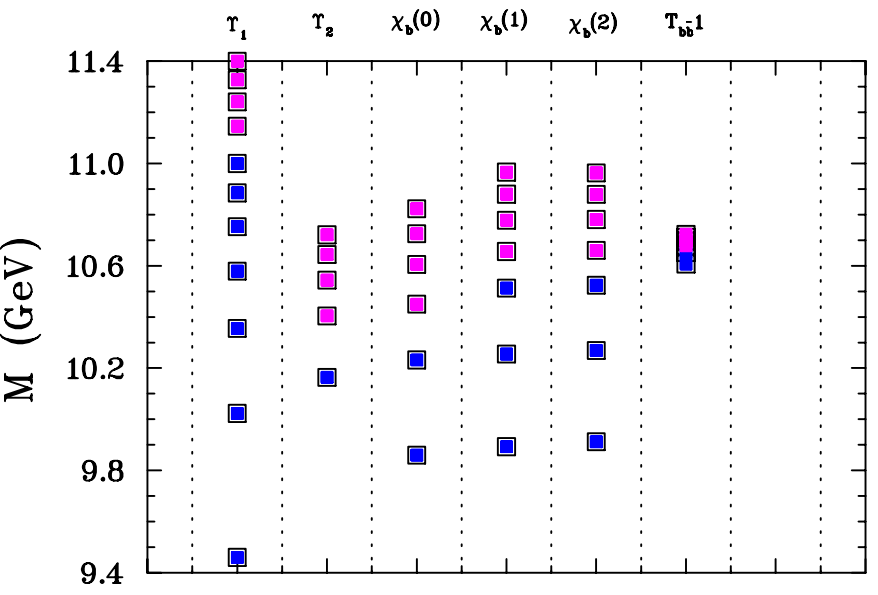} 
}

\centerline{\parbox{0.8\textwidth}{
\caption[] {\protect\small
Samples of spectral diagrams of meson excited-states sets analyzed are shown versus $J^{PC}$. Colors are used to display as follows: blue for PDG values~\cite{ParticleDataGroup:2024cfk} ($n = 3+$ states came from Part~I~\cite{Roper:2024ovj} and $n = 2$ states came from the current paper), green for missed states, and magenta for predicted states.
}
\label{fig:Money2} } }
\end{figure}

The analyzed spectra of the $\pi$, $\eta$, $\rho$, $\omega$, $\phi$, $f$, $a$, $\psi$, and $\Upsilon$ meson families are shown in Figure~\ref{fig:Money2}.
%--------------------------------------------
%------------------------------------------------------

%---------------------------------------------
\section{LHCb Exotics}

\begin{figure}[htb!]
%\vspace{-0.3cm}
\centering
{
    \includegraphics[width=0.50\textwidth,keepaspectratio]{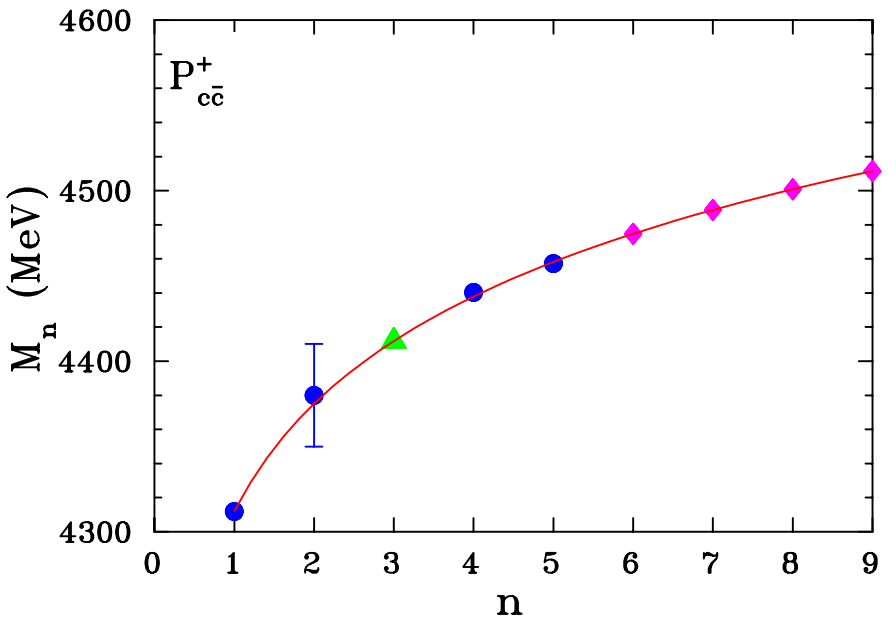}~~~ 
    \includegraphics[width=0.42\textwidth,keepaspectratio]{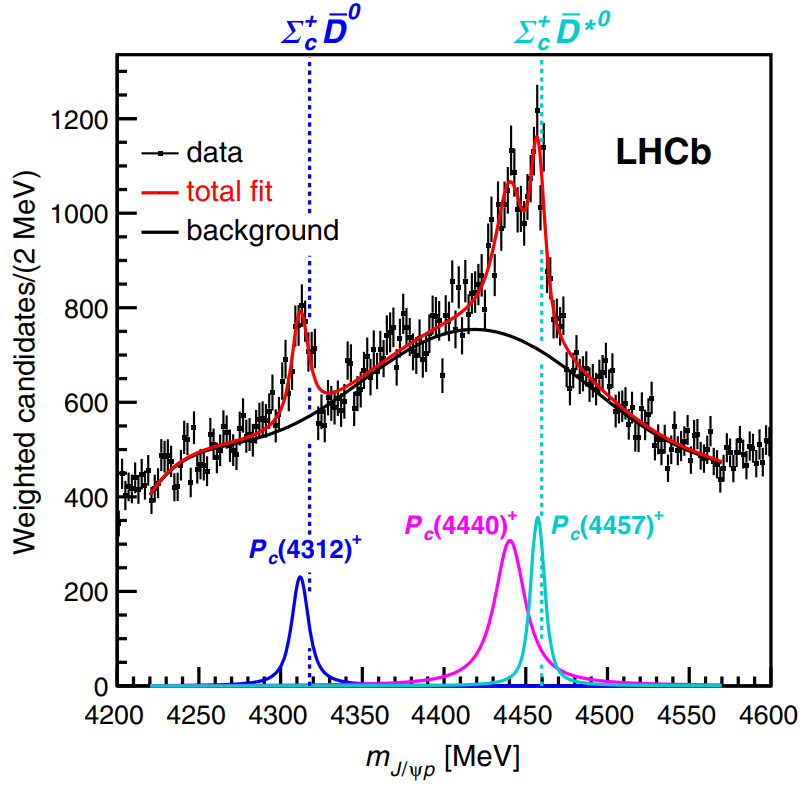}
}

\centerline{\parbox{0.8\textwidth}{
\caption[] {\protect\small
\underline{Left}: Data for $P_{c\bar{c}}^+(?^?)$ (blue circles): $P_{c\bar{c}}(4312)^+$, $P_{c\bar{c}}(4380)^+$, $P_{c\bar{c}}(4440)^+$, and $P_{c\bar{c}}(4457)^+$~\cite{ParticleDataGroup:2024cfk}. The green triangle is the calculated mass of the missing $P_{c\bar{c}}(4412)^+$ state with mass of $4411.8\pm 3.7~\mathrm{MeV}$. This state is weakly visitable but LHCb does not report it. Predicted states (magenta diamonds): $P_{c\bar{c}}(4475)^+$, $P_{c\bar{c}}(4489)^+$, $P_{c\bar{c}}(4501)^+$, and $P_{c\bar{c}}(4511)^+$ with masses of $4474.7\pm 3.4~\mathrm{MeV}$, $4488.7\pm 3.3~\mathrm{MeV}$, $4500.8\pm 3.3~\mathrm{MeV}$, and $4511.5\pm 3.3~\mathrm{MeV}$, respectively. The solid red curve presents the best-fit~\cite{Roper:2024ovj}.
\underline{Right}: Fit to the $\cos\theta_{Pc\bar{c}}$-weighted $m_{J\psi p}$ distribution with three BW amplitudes and a sixth-order polynomial background.
This fit is used to determine the central values of the masses and widths of the $P^+_{c\bar{c}}$ states. The mass thresholds for the $\Sigma_c^+\bar{D}^0$ and $\Sigma_c^+\bar{D}^{\ast 0}$ final states are superimposed~\cite{LHCb:2019kea}.
}
\label{fig:lhcb} } }
\end{figure}

QCD gives rise to Hadron Spectrum~\cite{Gell-Mann:1964ewy, Zweig:1964jf}. PDG2024~\cite{ParticleDataGroup:2024cfk} reports that many $q\bar{q}$ and $qqq$ states have been observed - more than 200 and 100 states, respectively. In addition, $q\bar{q}q\bar{q}$ and $qqq\bar{q}q$ are not forbidden, or we do not know it yet. Recently, LHCb Collaboration reported four positively charged exotic states $P^+_{c\bar{c}}$ with $I(J^P) = 1/2(?^?)$ and significance $5.4-7.3\sigma$~\cite{LHCb:2015yax, LHCb:2019kea}. LHCb found them in the reaction $\Lambda^0_b \to P^+_{c\bar{c}}K^- \to (J/\psi p)K^-$. As our Part~I~\cite{Roper:2024ovj} reported (see Fig.~15), one missed state with the mass of $4411.8\pm 3.7~\mathrm{MeV}$ and four predicted states with masses of $4474.7\pm 3.4~\mathrm{MeV}$, $4488.7\pm 3.3~\mathrm{MeV}$, $4500.8\pm 3.3~\mathrm{MeV}$, and $4511.5\pm 3.3~\mathrm{MeV}$. One can compare it with the LHCb spectrum shown in Fig.~6~\cite{LHCb:2019kea}.  Obviously, our predicted states and LHCb spectrum indicate that there is room for future study (Fig.~\ref{fig:lhcb}).
%--------------------------------------------

%---------------------------------------------------
\section{Measurability of Predicted Excited States}
The four predicted excited states of the duo sets that are analyzed in this article, because of their inherent widths compared to their mass differences, may be too close together for their masses to be measurable in experiments. There is no way to predict the widths; we assign a width to the predicted states of a set equal to the average width of the known states of the 
set. Two examples ($N1/2^+$ and $a_0$) are shown below of duo sets whose predicted states' masses are unlikely to be measurable for $N1/2^+$ and possibly measurable for $a_0$ (Fig.~\ref{fig:N1/2+MW}) and one ($\Upsilon$) whose masses are likely to be measurable (Fig.~\ref{fig:UpsMW}).
The BW resonance with width curves is calculated using the normalized relativistic accurate Equation~(\ref{eq:eqa}) given in ``\textit{A Brief History of Mass}.''~\cite{Wallen:2025brk}: 
\begin{equation}
    P(s,M,\Gamma)=\frac{\Gamma}{2\pi}\frac{4M^2+\Gamma^2}{(s^2-M^2+\Gamma^2/4)^2+M^2\Gamma^2}
    , \>
\label{eq:eqa}
\end{equation}

%---------------------------------------------------------
\subsection{$N1/2^+$ Baryon and $a_0(980)$ Meson Excited States with Widths}
\begin{figure}
    \centering
    \includegraphics[width=0.5\linewidth]{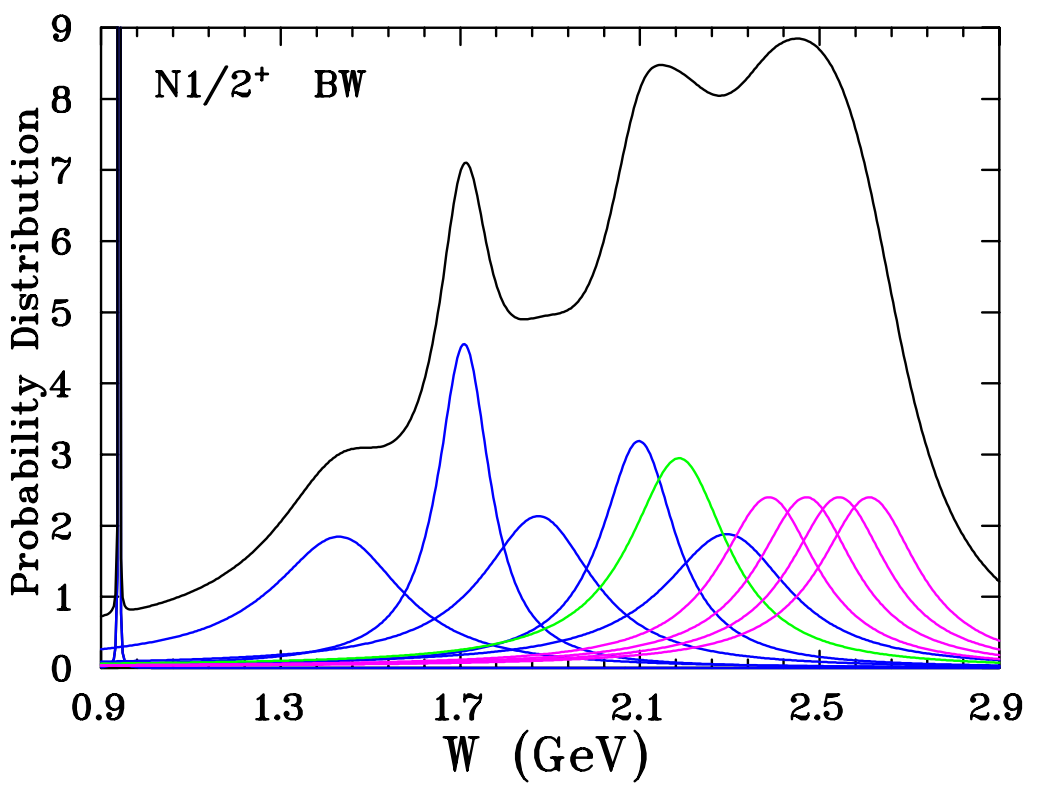}~~~
    \includegraphics[width=0.5\linewidth]{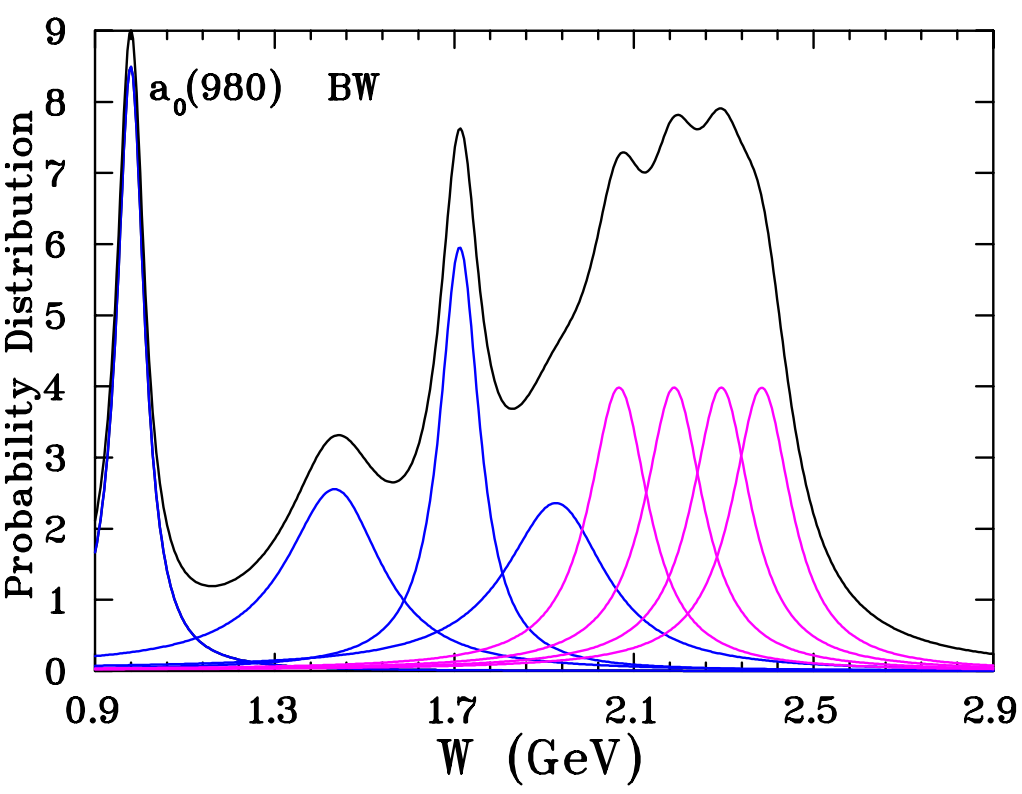}
    \caption{Samples for the ``bump hunting'' in the baryon ($N1/2^+$) and meson ($a_0(980)$) cases.
    Blue BW curves show PDG2024 values, the $N1/2^+$ green curve represents a missed state, and magenta curves show predicted states. Finally, black curves show the sum of all states' curves.
}
    \label{fig:N1/2+MW}
\end{figure}

The Roper-Strakovsky Universal Mass Equation (UME) fit to the $N1/2^+$ ($a_0(980)$) excited-state masses in Ref.~\cite{Roper:2024ovj} has PDG2024 values (blue), a missing state (green) inserted at $2.191-\mathrm{GeV}$ and predicts four higher-mass states at 2.391, 2.474, 2.547, and 
$2.614~\mathrm{GeV}$ (2.068, 2.191, 2.296, and $2.386~\mathrm{GeV}$). Assume that the missing state and the four predicted states have the average width $0.266~\mathrm{GeV}$ ($0.160~\mathrm{GeV}$) of the five known excited states above the neutron. See Fig.~\ref{fig:N1/2+MW} (left) for $N1/2^+$ and (right) for $a_0(980)$. It illustrates that the ``bump hunting'' technology is problematic to look for the predicted states. because of the overlapping resonances.

%---------------------------------------------------------
\subsection{$\Upsilon$ Meson Excited States with Widths}
\begin{figure}
    \centering
    \includegraphics[width=0.6\linewidth]{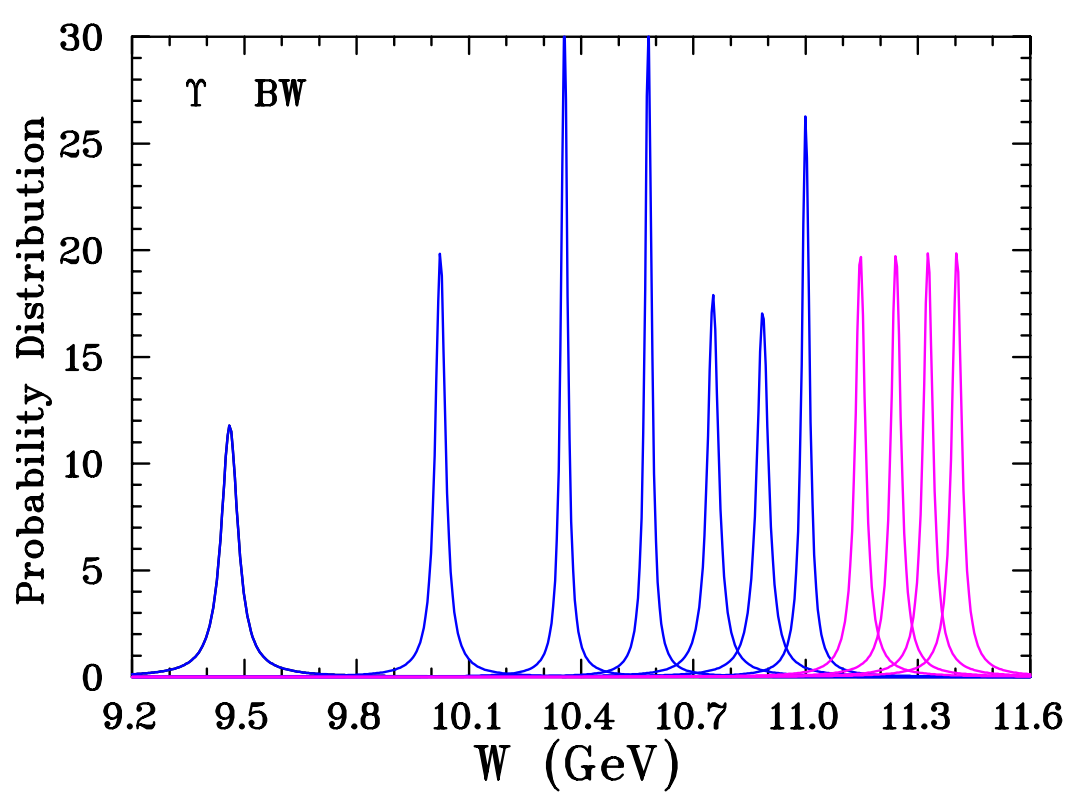}
    \caption{Sample for the ``bump hunting'' in the heavy vector meson ($\Upsilon$) case.
    Blue BW curves show PDG2024 values and magenta curves show predicted states.
}
    \label{fig:UpsMW}
\end{figure}

The Roper-Strakovsky UME fits to the $\Upsilon$ excited-state masses in Ref.~\cite{Roper:2024ovj} predict four higher-mass states at 11.146, 11.241, 11.327, and $11.404~\mathrm{GeV}$. Assume that the four predicted states have the average width ($0.0319~\mathrm{GeV}$) of the seven known excited states. See Fig.~\ref{fig:UpsMW}. It illustrates that the ``bump hunting'' technology is realistic to look for the predicted states, because the overlapping of the curves is small.

%---------------------------------------------
\section{Approximations}
After completing the UME fits to the mass sets above, we saw ways to approximate higher mass states for sets with only one mass in a set, similar to the $\Upsilon_2$ calculation.

%--------------------------
\subsection{Baryon Power Equation}
\begin{figure}
    \centering
    \includegraphics[width=0.6\linewidth]{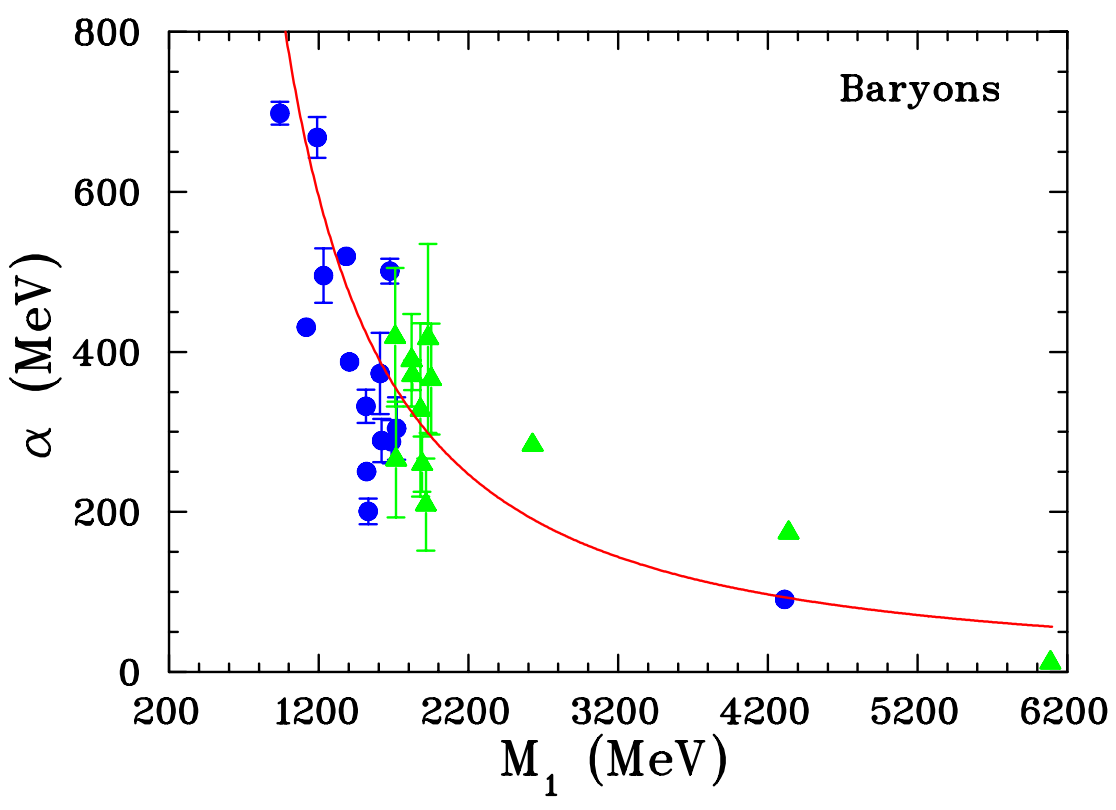}
    \caption{Baryon $\alpha$ vs $M_1$ power-equation fit [($\alpha=1.688\times10^7)M_1^{--1.446}~\mathrm{MeV}$]. Blue filled circles are from Part~I baryon sets (Table~1), green filled triangles are from this Part~II baryon sets (Table~6), and red curve is Parts~I and II fit to the power equation.}
    \label{fig:AlPwLw}
\end{figure}

A plot of fit parameter $\alpha$ versus $M_1$ for all baryon sets in Part~I and Part~II roughly fits a power equation, except for three Part~II baryon sets: $\Delta7/2^+$, $\Delta5/2^-$, and $\Lambda_c3/2^-$, which have exceptionally high $\alpha$ values. The recalculation for those three sets, with lower $\alpha$ values due to an assumed missing $M_2$, is included here rather than the original calculations. The power-equation [($\alpha=1.688\times10^7)M_1^{-1.446}~\mathrm{MeV}$] fit with the three lower $\alpha$ values is shown in Fig.~\ref{fig:AlPwLw}.

%--------------------------
\subsection{$\alpha$ Estimate for $\Upsilon_2$ Using $b\bar{b}$ Line Fit}
\begin{figure}[htb!]
%\vspace{-0.3cm}
\centering
{
    \includegraphics[width=0.5\textwidth,keepaspectratio]{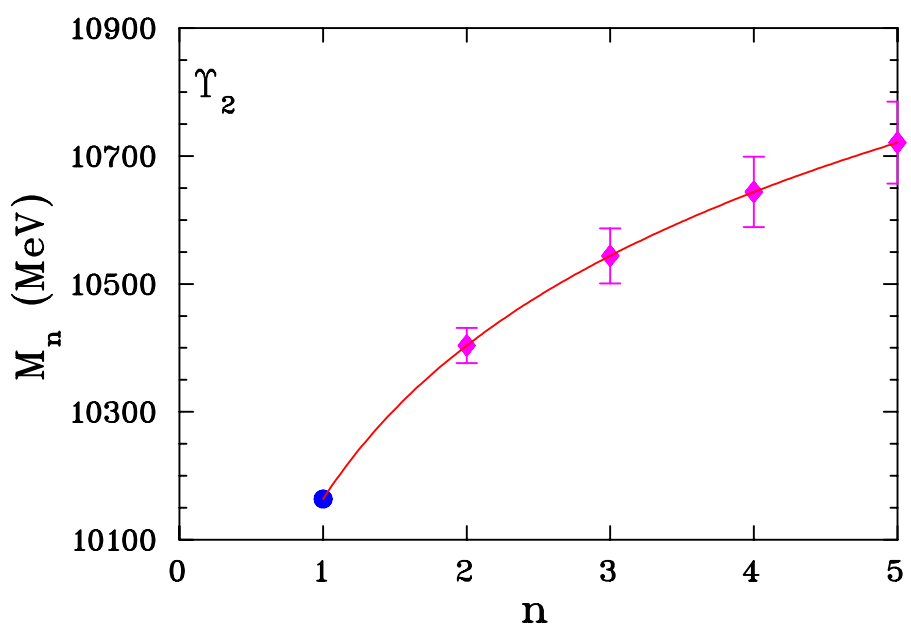} 
}

\centerline{\parbox{0.8\textwidth}{
\caption[] {\protect\small
Data for $\Upsilon_2(2^{-~-})$ (blue circle): 
$\Upsilon_2(10164)$~\cite{ParticleDataGroup:2024cfk}.
Predicted states generated by Eq.~(\ref{eq:eqbb}) (magenta diamonds): $\Upsilon_2(10404)$, $\Upsilon_2(10544)$, $\Upsilon_2(10644)$,  and $\Upsilon_2(10721)$, with masses of  $10404\pm 27~\mathrm{MeV}$, $10544\pm 43~\mathrm{MeV}$, $10644\pm 55~\mathrm{MeV}$, and $10721\pm 64~\mathrm{MeV}$, respectively.
The solid red curve presents the best-fit.
The fit parameter generated by Eq.~(\ref{eq:eqbbA}) is $\alpha = 346.2\pm 39.5~\mathrm{MeV}$.
}
\label{fig:Ups2} } }
\end{figure}

In Fig.~\ref{fig:Ups2}, a single $b\bar{b}~\Upsilon_2 ~0^-(2^{-~-})$ state with mass $10163.7\pm 1.4~\mathrm{MeV}$ is shown. Inserting this ground-state mass into Eq.~(\ref{eq:eqbb}) yields masses of higher-predicted states, shown in Fig.~\ref{fig:Ups2}.
%--------------------------------------------

%---------------------------------------------------------------
\subsection{$\psi(1^{-~-})$ Excited-States}

Our Part~I~\cite{Roper:2024ovj} reported a logarithmic fit for a series of excited $\psi$-meson states that are recorded in the Particle Data Listings~\cite{ParticleDataGroup:2024cfk}. $\psi$: $I^G(J^{PC}) = 0^-(1^{-~-})$, $c\bar{c}$ (Fig.~36).
We found that the state $\psi(2S)$ with a mass of $3686.097\pm 0.011~\mathrm{MeV}$ is well determined. Our recent analysis~\cite{Roper:2024ovj} flagged it. Now we redid the analysis to take into account this state.

\begin{figure}[htb!]
%\vspace{-0.3cm}
\centering
{
    \includegraphics[width=0.5\textwidth,keepaspectratio]{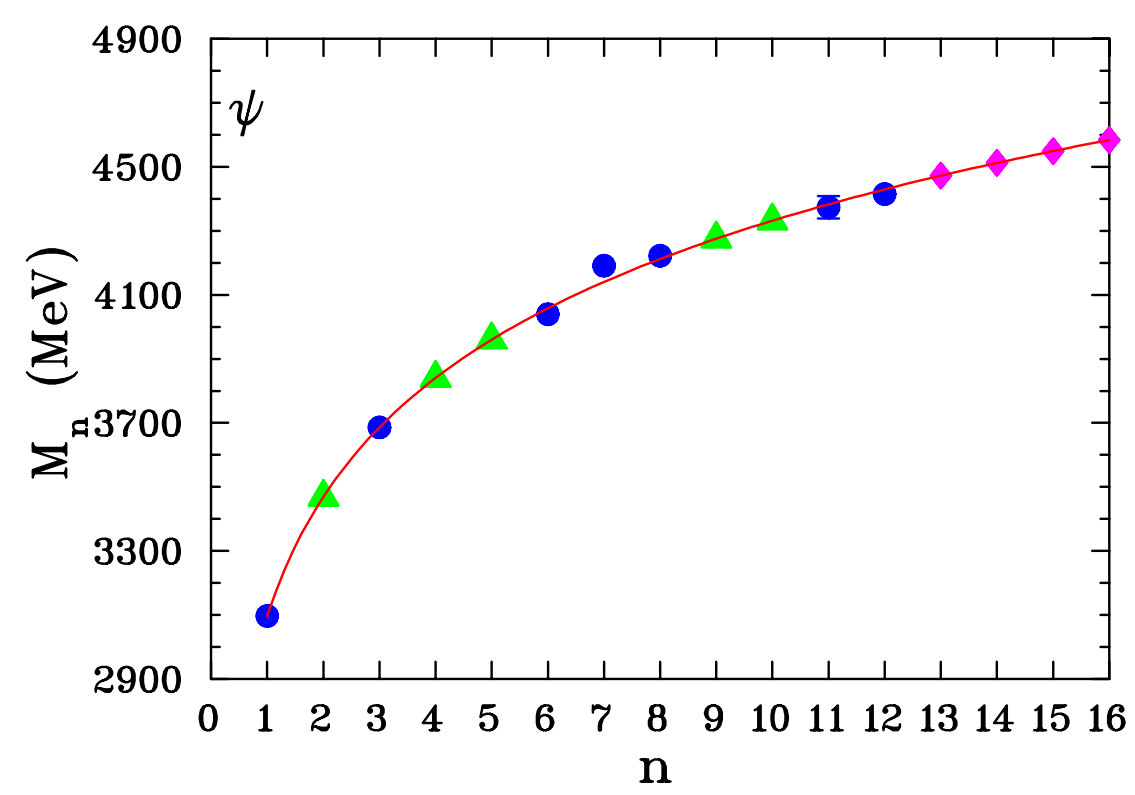} 
}

\centerline{\parbox{0.8\textwidth}{
\caption[] {\protect\small
Data for $\psi(1^{-~-})$ (blue circles): $J/\psi(3097)$, $\psi(3686)$, $\psi(4040)$, $\psi(4191)$, $\psi(4374)$, $\psi(4374)$, and $\psi(4415)$~\cite{ParticleDataGroup:2024cfk}. 
The green triangles are the calculated masses of the missing states $\psi(3469)$, $\psi(3840)$, $\psi(3960)$, $\psi(4275)$, and $\psi(4332)$ with masses of $3468.6\pm 0.1~\mathrm{MeV}$, $3840.4\pm 0.1~\mathrm{MeV}$, $3960.1\pm 0.1~\mathrm{MeV}$, $4275.3\pm 0.1~\mathrm{MeV}$, and $4331.8\pm 0.1~\mathrm{MeV}$, respectively.
Predicted states (magenta diamonds): $\psi(4472)$, $\psi(4512)$, $\psi(4549)$, and $\psi(4584)$ with masses of $4472.5\pm 0.1~\mathrm{MeV}$, $4512.2\pm 0.1~\mathrm{MeV}$, $4549.2\pm 0.1~\mathrm{MeV}$, and $4583.9\pm 0.1~\mathrm{MeV}$, respectively.
The solid red curve presents the best-fit.
The fit parameters are $\alpha = 531.3\pm 0.1~\mathrm{MeV}$ and $\beta = 3096.90\pm 0.03~\mathrm{MeV}$.
$\chi^2/\mathrm{DoF} = 1.2$ and CL = 31.1\%, taking into account the fact that we increased uncertainties of input data by a factor of 5.
}
\label{fig:fig36} } }
\end{figure}

The logarithmic fit to the BW masses (MeV) of the seven known $\psi$-meson excited states (blue circles) and four projected higher excited states (magenta diamonds) is shown in Fig.~\ref{fig:fig36}. $\psi$ has two states (instead of three in our previous analysis) reported in the PDG with masses $\psi(3770)$ and $\psi(4641)$ that do not fit 
Eq.~(\ref{eq:eq1}). We conclude that this mass is not well determined.
PDG gives very small uncertainties for heavy-mass mesons of approximately 0.2\% that do not allow Eq.~(\ref{eq:eq1}) to fit them well. We increased the uncertainties of the input data by a factor of 5 to present the results.
In addition, the five missing states (green triangles) are shown as calculated.
%--------------------------------------------
\section{Conclusion}
\raggedright
%--------------------------------------------
\begin{table}[htb!]

\centering \protect\caption{{Summary for free parameters $\alpha$ and $\beta$ of Eq.~(\ref{eq:eq2}) corresponding to the best-fit results for the 12 baryon excited-states sets. The PDG's BW mass estimation is within uncertainty of the obtained parameter $\beta$.}}

\vspace{2mm}
{%
\begin{tabular}{|c|c|c|}
\hline
Baryon           &  $\alpha$      &  $\beta$ \tabularnewline
                 & (MeV)          & (MeV)   \tabularnewline
\hline
$\Xi_b 3/2^-$    &  11.7$\pm$0.7  & 6087 \tabularnewline
$P_{c\bar{c}s}^01/2^-$& 174.0$\pm$5.9& 4338 \tabularnewline
$\Sigma 5/2^+$   & 209.2$\pm$57.7 & 1915 \tabularnewline
$\Lambda 3/2^+$  & 259.7$\pm$34.6 & 1890 \tabularnewline
$\Delta 1/2^+$   & 265.5$\pm$72.1 & 1716 \tabularnewline
$\Lambda_c 3/2^-$& 283.6$\pm$1.3  & 2628 \tabularnewline
$\Delta 5/2^+$   & 327.5$\pm$108.2& 1880 \tabularnewline
$\Delta 5/2^-$   & 365.9$\pm$69.2 & 1950 \tabularnewline
$\Lambda 5/2^-$  & 370.8$\pm$18.7 & 1825 \tabularnewline
$\Lambda 5/2^+$  & 389.5$\pm$57.7 & 1820 \tabularnewline
$\Delta 7/2^+$   & 415.9$\pm$72.8 & 1930 \tabularnewline
$\Delta 3/2^-$   & 418.4$\pm$86.6 & 1710 \tabularnewline
\hline
\end{tabular}} \label{tbl:tab3}
\end{table}
%--------------------------------------------
\begin{table}[htb!]

\centering \protect\caption{{Summary for free parameters $\alpha$ and $\beta$ of Eq.~(\ref{eq:eq2}) corresponding to the best-fit results for the 16 meson excited-states sets. The PDG's BW mass estimation is within uncertainty of the obtained parameter $\beta$.}}

\vspace{2mm}

{%
\begin{tabular}{|c|c|c|}
\hline
Meson            &  $\alpha$      &  $\beta$    \tabularnewline
                 & (MeV)          & (MeV)   \tabularnewline
\hline
$T_{b\bar{b}1}(1^{+-})$& 64.9$\pm$2.2& 10607 \tabularnewline
$D_{s1}(1^+)$    & 109.1$\pm$0.1  & 2459 \tabularnewline
$D_{s1}^\ast(1^-)$& 209.2$\pm$38.9& 2714 \tabularnewline
$D_1^\ast(1^-)$  & 222.2$\pm$31.7 & 2627 \tabularnewline
$T_{c\bar{c}\bar{s}1}(1^+)$& 318.8$\pm$63.5 & 3995 \tabularnewline
$\eta_2(2^{++})$ & 324.6$\pm$11.5 & 1617 \tabularnewline
$f_4(4^{++})$    & 435.7$\pm$86.6 & 2018 \tabularnewline
$h_b(1^{+-})$    & 520.1$\pm$1.7  & 9899 \tabularnewline
$\chi_{c2}(2^{++})$& 527.9$\pm$2.6& 3556 \tabularnewline
$\chi_{b0}(0^{++})$& 537.5$\pm$0.9& 9860 \tabularnewline
$a_2(2^{++})$    & 559.5$\pm$20.2 & 1318 \tabularnewline
$a_1(1^{++})$    & 613.1$\pm$23.1 & 1230 \tabularnewline
%$K_4^\ast(4^+)$  & 637.7$\pm$28.8 & 2048 \tabularnewline
$K_2^\ast(2^+)$  & 811.8$\pm$79.3 & 1427 \tabularnewline
$B_c(0^-)$       & 860.9$\pm$1.4  & 6274 \tabularnewline
$\eta_b(0^{-+})$ & 866.0$\pm$6.1  & 9299 \tabularnewline
$\eta_c(0^{-+})$ & 986.2$\pm$1.3  & 2984 \tabularnewline
\hline
\end{tabular}} \label{tbl:tab4}
\end{table}
%-------------------------------------------------------------------------

Twelve baryon and sixteen meson excited-states data sets that contain  only two lowest known masses (duo sets) are used, with Eq.~(\ref{eq:eq1}), to estimate, with uncertainties, the next four higher mass states in each set. The most accurate predictions, with very small uncertainty bars hidden by the data points, are for the $c$-quark sets [$D_{s1}$, $\Lambda_c3/2^-$, and $\chi_{c2}$] and the $b$-quark sets [$B_c$, $\eta_b$, $h_b$, and $\chi_{b0}$]. Of the remaining twenty-two sets of baryon and meson excited  states, the most accurate predictions are for the sets of states $ K^\ast_2$, $f_4$, $\eta_2$, $a_1$, $a_2$, and $\Lambda 5/2^-$.

It is amazing that equation $M_n = \alpha~Ln(n) + \beta$, can be used to accurately calculate the masses of several higher-state masses in an equal-quantum excited-state set using the first two measured masses to determine the parameters $\alpha$ and $\beta$:  $\beta = M_1$ and $\alpha = [(M_2 - M_1] / Ln(2)$. Of course, the accuracy of the predicted masses depends on the accuracy of the two measured masses. This should be of use to experimentalists.

After the two lowest-mass excited states in equal-quantum excited-states sets are reported at high accuracy from experiments, our universal mass equation for equal-quantum excited-states sets can be used to accurately predict several higher-mass states in the set. 

It is not surprising that baryons and mesons look similar because a baryon can be considered as a meson ($qq$) plus one more $q$. For light quarks at large $n$ from quasi-classics, one expects $\Delta M(n)\propto 1/n$. Then one can see that logarithm behavior gives stronger behavior than the quasi-classical case, and it works for heavy quarks as well.

Theory needs to do the following:

1. Produce the logarithm curve for equal-quantum excited-state sets (Eq.~(\ref{eq:eq1})).

2. Produce the linear relationship between $\alpha(M_1)$ and $M_1$ (Eq.~(5) for $b\bar{b}$ equal-quantum excited states sets.

%------------------------------------------------------------
\section*{Acknowledgments}

This work was supported in part by the U.~S.~Department of Energy, Office of Science, Office of Nuclear Physics, under Award No.~DE--SC0016583.

Both authors dedicate this document to the memory of their excellent colleague and friend Richard Allen Arndt.

%--------------------------------------------
%\begin{table}[htb!]
%
%\centering \protect\caption{{Vector Mesons, $I^G(J^{PC}) = 0^-(1^{-~-})$. 
%PDG masses are from \cite{ParticleDataGroup:2024cfk}}}
%
%\vspace{2mm}
%
%{%
%\begin{tabular}{|c|c|c|c|}
%\hline
%Meson      & Quark      & Mass  & $\alpha$ \tabularnewline
%           & Contents   & (MeV) & (MeV)    \tabularnewline
%\hline
%$\omega(782)$ & $\frac{u\bar{u}+d\bar{d}}{\sqrt{2}}$& 782.66   & 811.3 \tabularnewline
%$\phi(1020)$  & $s\bar{s}$                          & 1019.461 & 588.5\tabularnewline
%$J/\psi(1S)$  & $c\bar{c}$                          & 3096.900 & 617.3 \tabularnewline
%$\Upsilon(1S)$& $b\bar{b}$                          & 9460.40  & 810.6 \tabularnewline
%\hline
%\end{tabular}} \label{tbl:tabv}
%\end{table}
%-------------------------------------------------------------------------

%---------------------- REFERENCES ----------------------


\begin{thebibliography}{99}
\bibitem{Roper:2024ovj}
    L.~D.~Roper and I.~Strakovsky,
    ``Universal Mass Equation for Equal-Quantum Excited-States Sets~I,''
    Eur.\ Phys.\ J.\ A\ \textbf{61}, 102 (2025).
\bibitem{ParticleDataGroup:2024cfk}
    S.~Navas \textit{et al.} [Particle Data Group],
    ``Review of particle physics,''
    Phys.\ Rev.\ D\ \textbf{110}, 030001 (2024).
\bibitem{Amsler:2024}
    C.~Amsler \textit{et al} ``8.~Naming Scheme for Hadrons,'' in:
    S.~Navas \textit{et al.} [Particle Data Group],
    ``Review of particle physics,''
    Phys.\ Rev.\ D\ \textbf{110}, 030001 (2024).
\bibitem{Gell-Mann:1964ewy}
    M.~Gell-Mann,
    ``A schematic model of baryons and mesons,''
    Phys.\ Lett.\ \textbf{8}, 214 (1964).
\bibitem{Zweig:1964jf}
    G.~Zweig,
    ``An SU(3) model for strong interaction symmetry and its breaking. Version 2,''
    Preprint CERN-TH-412, 1964.
\bibitem{LHCb:2015yax}
    R.~Aaij \textit{et al.} [LHCb],
    ``Observation of $J/\psi p$ resonances consistent with pentaquark states in $\Lambda_b^0 \to J/\psi K^- p$ decays,''
    Phys.\ Rev.\ Lett.\ \textbf{115}, 072001 (2015).
\bibitem{LHCb:2019kea}
    R.~Aaij \textit{et al.} [LHCb],
    ``Observation of a narrow pentaquark state, $P_c(4312)^+$, and of two-peak structure of the $P_c(4450)^+$,''
    Phys.\ Rev.\ Lett.\ \textbf{122}, 222001 (2019).
\bibitem{Wallen:2025brk}
    S.~Willenbrock, 
    ``A brief history of mass,'' 
    [arXiv:2503.07866 [hep-ph]].
%\bibitem{Eichten:1974af}
%    E.~Eichten, K.~Gottfried, T.~Kinoshita, J.~B.~Kogut, K.~D.~Lane, and T.~M.~Yan,
%    ``The spectrum of Charmonium,''
%    Phys.\ Rev.\ Lett.\ \textbf{34}, 369 (1975) 
%    [erratum: Phys.\ Rev.\ Lett.\ \textbf{36}, 1276 (1976)].
%\bibitem{Eichten:1978tg}
%    E.~Eichten, K.~Gottfried, T.~Kinoshita, K.~D.~Lane, and T.~M.~Yan,
%    ``Charmonium: The model,''
%    Phys.\ Rev.\ D\ \textbf{17}, 3090 (1978)
%    [erratum: Phys.\ Rev.\ D\ \textbf{21}, 313 (1980)].
%\bibitem{Brambilla:1999ja}
%    N.~Brambilla and A.~Vairo,
%    ``Quark confinement and the hadron spectrum,''
%    [arXiv:hep-ph/9904330 [hep-ph]].
%\bibitem{Yamamoto:2008jz}
%    A.~Yamamoto, H.~Suganuma, and H.~Iida,
%    ``Lattice QCD study of the heavy-heavy-light quark potential,''
%    Phys.\ Rev.\ D\ \textbf{78}, 014513 (2008).
\end{thebibliography}
\end{document}